\documentclass[10pt,twocolumn,english,10pt,english,twocolumn,aps,prd,longbibliography]{revtex4-1}
\usepackage{lmodern}
\usepackage[T1]{fontenc}
\usepackage[latin9]{inputenc}
\usepackage{babel}
\usepackage{amsmath}
\usepackage{graphicx}
\usepackage[unicode=true,
 bookmarks=false,
 breaklinks=false,pdfborder={0 0 1},colorlinks=false,colorlinks=true,
  urlcolor=blue,
  linkcolor=blue,
  citecolor=blue]{hyperref}

\makeatletter
\usepackage{lmodern}
\usepackage[T1]{fontenc}
\usepackage{prettyref}
\usepackage{textgreek}
\usepackage{inputenc}
\usepackage{slashed}
\usepackage{hyperref}

\makeatother

\begin{document}
\title{Numerical approach to the semiclassical method of pair production
for arbitrary spins and photon polarization}
\author{Tobias N. Wistisen}
\affiliation{Max-Planck-Institut f{\"u}r Kernphysik, Saupfercheckweg 1, D-69117 Heidelberg,
Germany}
\begin{abstract}
In this paper we show how to recast the results of the semiclassical
method of Baier, Katkov \& Strakhovenko for pair production, including
the possibility of specifying all the spin states and photon polarization,
in a form that is suitable for numerical implementation. In this case,
a new type of integral appears in addition to the ones required for
the radiation emission process. We compare the resulting formulas with those obtained for a short pulse plane wave external field by
using the Volkov state. We investigate the applicability of
the local constant field approximation for the proposed upcoming experiments
at FACET II at SLAC and LUXE at DESY. Finally, we provide results on the dependence
of the pair production rate on the relative polarization between a
linearly polarized laser pulse and a linearly polarized incoming high
energy photon. We observe that even in the somewhat intermediate intensity
regime of these experiments, there is roughly a factor of $2$ difference
between the pair production rates corresponding to the two relative photon polarizations, which is of interest in light of the vacuum birefringence of QED.
\end{abstract}
\maketitle

\section{Introduction}

In view of the rapid development of laser technology, the consideration
of nonlinear QED effects in the interaction of light with matter is
increasingly important. Examples of such processes is quantum radiation
emission, experimentally seen in nonlinear Compton scattering in \cite{bula1996observation}
and also in channeling radiation in crystals \citep{BAK1985491,BAK1988525,PhysRevLett.43.1723,1402-4896-24-3-015,ANDERSEN1982209,PhysRevB.31.68,PhysRevLett.42.1148,PhysRevD.86.072001,RevModPhys.77.1131,PhysRevLett.112.254801}.
Recently it has also been possible to see multi-photon emission in,
or close to, the quantum regime, the so-called quantum radiation reaction
studied extensively theoretically in e.g. \cite{PhysRevLett.105.220403,PhysRevLett.111.054802,PhysRevLett.112.015001,PhysRevLett.112.145003,ILDERTON2013481,PhysRevLett.116.044801,PhysRevLett.113.134801,PhysRevD.85.101701,PhysRevLett.110.070402,dinu2018single,PhysRevA.91.033415,dinu2019approximating}
and also recently studied experimentally in crystal and laser fields \cite{Wistisen2018experimental,PhysRevX.8.011020,PhysRevX.8.031004,Wistisen2019exp}.
Future experiments at SLAC, DESY and the Extreme Light Infrastructure,
aim to study these processes further into the nonlinear regime
\cite{LUXE,FACET,Turcu2016}. Another related nonlinear process of
strong-field QED is that of electron-positron pair production, for
the case of a laser field, called the nonlinear Breit-Wheeler process. This is the nonlinear counterpart of the Breit-Wheeler process \cite{PhysRev.46.1087} in the sense that absorption of several photons from
the strong field occurs. This has been studied theoretically in a
short pulse using the Volkov state in e.g. \cite{PhysRevA.86.052104,PhysRevA.87.042106,PhysRevLett.117.213201,Heinzl2010250,PhysRevLett.108.240406,PhysRevD.93.053011,NOUSCH2016162,PhysRevD.94.013010,PhysRevD.91.013009,PhysRevD.93.085028}, see also \cite{PhysRevLett.114.143201} for the effect of recollision in the pair production process.
This process is also the subject of the current paper, but with the
focus on how to treat this process in more general field configurations.
In particular we show how the semiclassical method of Baier, Katkov
and Strakhovenko \cite{baier1968processes,Baier1998} in its most
general form, including spins and polarizations, can be recast into
a form that is suitable for numerical implementation. The strength
of this approach is that it can be used in any background field, as
only the Lorentz force trajectory of the produced electron in this
field is required, which is easily found numerically. This stands
in contrast to the conventional Furry picture approach where wave
functions in the background field must be found. The scheme presented in this paper could also be useful for polarization and spin effect studies, such as the ones seen in 
\cite{PhysRevD.91.125026,PhysRevLett.123.174801,WAN2020135120}. The semiclassical
approach is an approximation, the limits of which are discussed by
the authors themselves and additionally in e.g. \cite{Wistisen2019,PhysRevA.98.022131},
with the main criterion being, that the notion of a classical trajectory
should be reasonable, or that the quantum numbers associated with the
motion should be large. 

Below, the relativistic metric $+---$ is employed. We will use Feynman
notation to write $\ensuremath{\slashed{a}=a_{\mu}\gamma^{\mu}}$,
where $\ensuremath{a^{\mu}}$ is a generic 4-vector, and we will use
the shorthand for the product of 4-vectors, $ab=a^{\mu}b_{\mu}$.
We will use units where $\hbar=c=1$, and $e$ is the
elementary charge ($e^{2}\sim1/137$).

\section{Semiclassical Pair production}

Baier et. al write the pair production probability in the semiclassical
formalism, in the most general form, as \cite{Baier1998}

\begin{equation}
dP=\frac{e^{2}}{(2\pi)^{2}\omega}\left|\int_{-\infty}^{\infty}R_{p}(t)e^{i\frac{\varepsilon_{-}}{\varepsilon_{+}}\omega\left(t-\boldsymbol{n}\cdot\boldsymbol{x}_{-}\right)}dt\right|^{2}d^{3}\boldsymbol{p}_{-}\label{eq:Baierformula}
\end{equation}
where $dP$ is the differential transition probability, $\omega$ the energy of the
incoming photon which converts to a pair, $\boldsymbol{n}$ is a unit
vector along the momentum of this photon such that $\boldsymbol{k}=\omega\boldsymbol{n}$,
$\boldsymbol{x}_{-}(t)$ is the trajectory of an electron that solves
the Lorentz force equation, $\boldsymbol{v}_{-}=d\boldsymbol{x}_{-}/dt$
and $\boldsymbol{p}_{-}=\varepsilon_{-}\boldsymbol{v}_{-}$, $\varepsilon_{-}$
being the electron energy and

\begin{equation}
R_{p}(t)=i\phi_{-}^{\dagger}\left(A(t)-i\boldsymbol{\sigma}\cdot\boldsymbol{B}(t)\right)\phi_{+},\label{eq:R}
\end{equation}
\begin{align}
A(t) & =N_p\varepsilon_{-}\omega\boldsymbol{v}_{-}\cdot\left(\boldsymbol{\epsilon}\times\boldsymbol{n}\right),\label{eq:A}
\end{align}

\begin{align}
\boldsymbol{B}(t) & =N_p\left[\boldsymbol{\epsilon}\left\{ \left(\varepsilon_{-}+m\right)\omega-\varepsilon_{-}\omega\boldsymbol{v}_{-}\cdot\boldsymbol{n}\right\} \right.\nonumber \\
 & \left.-2\varepsilon_{-}^{2}\boldsymbol{v}_{-}\left(\epsilon\cdot\boldsymbol{v}_{-}\right)+\varepsilon_{-}\omega\boldsymbol{n}\left(\boldsymbol{\epsilon}\cdot\boldsymbol{v}_{-}\right)\right],\label{eq:B}
\end{align}

\begin{equation}
N_p=\frac{1}{\sqrt{4\varepsilon_{+}\varepsilon_{-}\left(\varepsilon_{-}+m\right)\left(\varepsilon_{+}+m\right)}},\label{eq:N}
\end{equation}
$\varepsilon_{+}=\omega-\varepsilon_{-}$, $\phi_{-}$ and $\phi_{+}$
are the 2-component spinors of the electron and positron respectively and $\boldsymbol{\epsilon}$ is the polarization of the incoming photon.
We have here re-written $\boldsymbol{B}(t)$ as compared to that found
in \cite{Baier1998}, which was achieved by using that $\boldsymbol{p}_{-}^{2}=\varepsilon_{-}^{2}-m^{2}$.
Note here that if we choose the quantization axis along $z$ then
$\left(\begin{array}{cc}
1 & 0\end{array}\right)^{T}$ corresponds to the spin-up state for the electron, while $\left(\begin{array}{cc}
0 & 1\end{array}\right)^{T}$ is the spin-up state for the positron. The integration over $d^{3}\boldsymbol{p}_{-}$
should be understood as the asymptotic momentum of the trajectory
when the external field has gone to zero. Therefore one must find
a trajectory for each final momentum, whereas for radiation emission
one only needs the trajectory corresponding to the initial state,
and therefore the semiclassical approach is typically more numerically
demanding for pair production. Note that the above formula only requires the electron trajectory, which is an arbitrary choice made during the derivation, where the summation over the final states of the positron was carried out, instead of that over the electron. This means that in the semiclassical approach $\boldsymbol{p}_{+}(t)=\boldsymbol{k}-\boldsymbol{p}_{-}(t)$ \cite{Baier1998}.
In order to calculate the
integral from Eq. \eqref{eq:Baierformula} we need the integrals

\begin{equation}
\int\boldsymbol{v}_{-}e^{i\frac{\varepsilon_{-}}{\varepsilon_{+}}\omega\left(t-\boldsymbol{n}\cdot\boldsymbol{x}_{-}\right)}dt,\label{eq:vint}
\end{equation}

\begin{equation}
\int e^{i\frac{\varepsilon_{-}}{\varepsilon_{+}}\omega\left(t-\boldsymbol{n}\cdot\boldsymbol{x}_{-}\right)}dt,\label{eq:expint}
\end{equation}

\begin{equation}
\int\boldsymbol{v}_{-}\left(\boldsymbol{\epsilon}\cdot\boldsymbol{v}_{-}\right)e^{i\frac{\varepsilon_{-}}{\varepsilon_{+}}\omega\left(t-\boldsymbol{n}\cdot\boldsymbol{x}_{-}\right)}dt.\label{eq:newint}
\end{equation}
While the first two integrals of Eq. \eqref{eq:vint} and \eqref{eq:expint}
are also encountered in the radiation emission process as can be seen
in e.g. \cite{PhysRevD.100.116001,PhysRevD.90.125008,PhysRevD.92.045045},
the third integral of Eq. \eqref{eq:newint} does not, and therefore
we will need to rewrite this in a similar fashion as what is done
for the first two. This amounts to an integration by parts and removing
the boundary terms, such that the integrals with an integrand proportional
to acceleration are obtained. The justification for this, is that
terms related to what happens in the infinite past and future, where
the field has turned off, should not have an effect on the result.
As a sanity check, we will compare the results obtained when using
the Volkov state solution of the Dirac equation in the background
field, where we will see that the results are indistinguishable. We
are working in the limit where the electrons and positrons will be
ultra relativistic. We then define the quantities in analogy to the
radiation emission process as

\begin{align}
\boldsymbol{I} & =\int\frac{\boldsymbol{n}\times\left[\left(\boldsymbol{n}-\boldsymbol{v}_{-}\right)\times\dot{\boldsymbol{v}}_{-}\right]}{\left(1-\boldsymbol{n}\cdot\boldsymbol{v}_{-}\right)^{2}}e^{i\frac{\varepsilon_{-}}{\varepsilon_{+}}\omega\left(t-\boldsymbol{n}\cdot\boldsymbol{x}_{-}\right)}dt\nonumber \\
 & =\int\frac{d}{dt}\left[\frac{\boldsymbol{n}\times\left(\boldsymbol{n}\times\boldsymbol{v}_{-}\right)}{1-\boldsymbol{n}\cdot\boldsymbol{v}_{-}}\right]e^{i\frac{\varepsilon_{-}}{\varepsilon_{+}}\omega\left(t-\boldsymbol{n}\cdot\boldsymbol{x}_{-}\right)}dt\nonumber \\
 & =-i\frac{\varepsilon_{-}}{\varepsilon_{+}}\omega\int\boldsymbol{n}\times\left(\boldsymbol{n}\times\boldsymbol{v}_{-}\right)e^{i\frac{\varepsilon_{-}}{\varepsilon_{+}}\omega\left(t-\boldsymbol{n}\cdot\boldsymbol{x}_{-}\right)}dt\nonumber \\
 & \simeq-i\frac{\varepsilon_{-}}{\varepsilon_{+}}\omega\int\left(\boldsymbol{n}-\boldsymbol{v}_{-}\right)e^{i\frac{\varepsilon_{-}}{\varepsilon_{+}}\omega\left(t-\boldsymbol{n}\cdot\boldsymbol{x}_{-}\right)}dt,\label{eq:I}
\end{align}
where in the last line we have neglected terms suppressed by at least
$1/\gamma_-$, with $\gamma_-=\varepsilon_-/m$, compared to the dominant ones, and we have that

\begin{align}
 & \int e^{i\frac{\varepsilon_{-}}{\varepsilon_{+}}\omega\left(t-\boldsymbol{n}\cdot\boldsymbol{x}_{-}\right)}dt\nonumber \\
 & =\int\frac{1}{i\frac{\varepsilon_{-}}{\varepsilon_{+}}\omega\left(1-\boldsymbol{n}\cdot\boldsymbol{v}_{-}\right)}\frac{d}{dt}e^{i\frac{\varepsilon_{-}}{\varepsilon_{+}}\omega\left(t-\boldsymbol{n}\cdot\boldsymbol{x}_{-}\right)}dt\nonumber \\
 & =-\int e^{i\frac{\varepsilon_{-}}{\varepsilon_{+}}\omega\left(t-\boldsymbol{n}\cdot\boldsymbol{x}_{-}\right)}\frac{d}{dt}\frac{1}{i\frac{\varepsilon_{-}}{\varepsilon_{+}}\omega\left(1-\boldsymbol{n}\cdot\boldsymbol{v}_{-}\right)}dt\nonumber \\
 & =\frac{i}{\frac{\varepsilon_{-}}{\varepsilon_{+}}\omega}\int\frac{\boldsymbol{n}\cdot\dot{\boldsymbol{v}}_{-}}{\left(1-\boldsymbol{n}\cdot\boldsymbol{v}_{-}\right)^{2}}e^{i\frac{\varepsilon_{-}}{\varepsilon_{+}}\omega\left(t-\boldsymbol{n}\cdot\boldsymbol{x}_{-}\right)}dt\nonumber \\
 & =\frac{i}{\frac{\varepsilon_{-}}{\varepsilon_{+}}\omega}J,\label{eq:expintrelate}
\end{align}
where we have then defined

\begin{equation}
J=\int\frac{\boldsymbol{n}\cdot\dot{\boldsymbol{v}}_{-}}{\left(1-\boldsymbol{n}\cdot\boldsymbol{v}_{-}\right)^{2}}e^{i\frac{\varepsilon_{-}}{\varepsilon_{+}}\omega\left(t-\boldsymbol{n}\cdot\boldsymbol{x}_{-}\right)}dt.\label{eq:Jdefine}
\end{equation}
Then from Eq. \eqref{eq:I} and \eqref{eq:expintrelate} we have that

\begin{align}
 & \int\boldsymbol{v}_{-}e^{i\frac{\varepsilon_{-}}{\varepsilon_{+}}\omega\left(t-\boldsymbol{n}\cdot\boldsymbol{x}_{-}\right)}dt\nonumber \\
 & =\frac{i}{\frac{\varepsilon_{-}}{\varepsilon_{+}}\omega}\left[\boldsymbol{n}J-\boldsymbol{I}\right].\label{eq:vintrelate}
\end{align}
Now we follow the same approach for the new integral

\begin{align}
 & \int\boldsymbol{v}_{-}\left(\boldsymbol{\epsilon}\cdot\boldsymbol{v}_{-}\right)e^{i\frac{\varepsilon_{-}}{\varepsilon_{+}}\omega\left(t-\boldsymbol{n}\cdot\boldsymbol{x}_{-}\right)}dt\nonumber \\
 & =\int\frac{\boldsymbol{v}_{-}\left(\boldsymbol{\epsilon}\cdot\boldsymbol{v}_{-}\right)}{i\frac{\varepsilon_{-}}{\varepsilon_{+}}\omega\left(1-\boldsymbol{n}\cdot\boldsymbol{v}_{-}\right)}\frac{d}{dt}e^{i\frac{\varepsilon_{-}}{\varepsilon_{+}}\omega\left(t-\boldsymbol{n}\cdot\boldsymbol{x}_{-}\right)}dt\nonumber \\
 & =-\int\frac{d}{dt}\left[\frac{\boldsymbol{v}_{-}\left(\boldsymbol{\epsilon}\cdot\boldsymbol{v}_{-}\right)}{i\frac{\varepsilon_{-}}{\varepsilon_{+}}\omega\left(1-\boldsymbol{n}\cdot\boldsymbol{v}_{-}\right)}\right]e^{i\frac{\varepsilon_{-}}{\varepsilon_{+}}\omega\left(t-\boldsymbol{n}\cdot\boldsymbol{x}_{-}\right)}dt\nonumber \\
 & =\frac{i}{\frac{\varepsilon_{-}}{\varepsilon_{+}}\omega}\boldsymbol{K},\label{eq:newint-1}
\end{align}
with

\begin{align}
\boldsymbol{K} & =\int\left(\frac{\left[\boldsymbol{a}_{-}\left(\boldsymbol{\epsilon}\cdot\boldsymbol{v}_{-}\right)+\boldsymbol{v}_{-}\left(\boldsymbol{\epsilon}\cdot\boldsymbol{a}_{-}\right)\right]}{1-\boldsymbol{n}\cdot\boldsymbol{v}_{-}}\right.\nonumber \\
 & \left.+\frac{\boldsymbol{v}_{-}\left(\boldsymbol{\epsilon}\cdot\boldsymbol{v}_{-}\right)\left(\boldsymbol{n}\cdot\boldsymbol{a}_{-}\right)}{\left(1-\boldsymbol{n}\cdot\boldsymbol{v}_{-}\right)^{2}}\right)e^{i\frac{\varepsilon_{-}}{\varepsilon_{+}}\omega\left(t-\boldsymbol{n}\cdot\boldsymbol{x}_{-}\right)}dt.\label{eq:K}
\end{align}
Then we may rewrite the integrals involving $A(t)$ and $\boldsymbol{B}(t)$
in the following fashion

\begin{align}
 & \int_{-\infty}^{\infty}A(t)e^{i\frac{\varepsilon_{-}}{\varepsilon_{+}}\omega\left(t-\boldsymbol{n}\cdot\boldsymbol{x}\right)}dt\nonumber \\
 & =\frac{i}{\frac{\varepsilon_{-}}{\varepsilon_{+}}\omega}\varepsilon_{-}\omega N\left[\boldsymbol{n}J-\boldsymbol{I}\right]\cdot\left(\boldsymbol{\epsilon}\times\boldsymbol{n}\right),\label{eq:Arewrite}
\end{align}

\begin{align}
 & \int_{-\infty}^{\infty}\boldsymbol{B}(t)e^{i\frac{\varepsilon_{-}}{\varepsilon_{+}}\omega\left(t-\boldsymbol{n}\cdot\boldsymbol{x}\right)}dt\nonumber \\
 & =\frac{i}{\frac{\varepsilon_{-}}{\varepsilon_{+}}\omega}N\left[\boldsymbol{\epsilon}\left\{ \left(\varepsilon_{-}+m\right)\omega J-\varepsilon_{-}\omega\left[\boldsymbol{n}J-\boldsymbol{I}\right]\cdot\boldsymbol{n}\right\} \right.\nonumber \\
 & \left.-2\varepsilon_{-}^{2}\boldsymbol{K}+\varepsilon_{-}\omega\boldsymbol{n}\left(\boldsymbol{\epsilon}\cdot\left[\boldsymbol{n}J-\boldsymbol{I}\right]\right)\right]\nonumber \\
 & =\frac{i}{\frac{\varepsilon_{-}}{\varepsilon_{+}}\omega}N\left[\boldsymbol{\epsilon}\omega mJ-2\varepsilon_{-}^{2}\boldsymbol{K}-\varepsilon_{-}\omega\boldsymbol{n}\left(\boldsymbol{\epsilon}\cdot\boldsymbol{I}\right)\right],\label{eq:Brewrite}
\end{align}
where we used that $\boldsymbol{\epsilon}\cdot\boldsymbol{n}=0$ and
that $\boldsymbol{I}\cdot\boldsymbol{n}=0$ as can be seen from Eq.
\eqref{eq:I}. Now one is able to perform the computation. One must
simply calculate the $\boldsymbol{I}$, $J$ and $\boldsymbol{K}$
integrals numerically based on the trajectory which can be obtained
by solving the Lorentz force equation, but where we recommend to follow
the approach developed in \cite{PhysRevD.90.125008} to deal with cancellations
between large terms as seen in e.g. $1-\boldsymbol{n}\cdot\boldsymbol{v}_{-}$
as $\boldsymbol{n}\cdot\boldsymbol{v}_{-}$ is close to $1$. One
can then pick the spin and polarization states and calculate the transition
probability for each combination. It therefore consumes nearly the
same computational resources to keep all the information regarding
spin and polarization, as it is the computation of $\boldsymbol{I}$,
$J$ and $\boldsymbol{K}$ which is demanding and here only $\boldsymbol{K}$
depends on the photon polarization (but not on the spins of the electron and positron).

\section{Volkov state approach}

The Dirac equation in a background field, given by the 4-vector potential
$A^{\mu}$

\begin{equation}
\left(i\slashed{\partial}+e\slashed{A}-m\right)\psi=0,
\end{equation}
has an exact solution when $A^{\mu}$ is a plane wave, i.e. it depends
on space-time only through the variable $\varphi=k_{0}x$ where $k_{0}$
is the wave vector characterizing the plane wave, with $k_{0}^2=0$. In this case the
electron solution is given by

\begin{equation}
\psi_{-}(x)=\frac{1}{\sqrt{2\varepsilon_{-}}}\left(1-\frac{e\slashed{k}_{0}\slashed{A}}{2k_{0}p_{-}}\right)ue^{iS_{-}},\label{eq:volkovstate}
\end{equation}
where $p$ is the asymptotic 4-momentum of the electron, (we have
set the quantization volume equal to $1$), $u$ is the free particle electron
bispinor which is reached asymptotically and where

\begin{equation}
S_{-}=-p_{-}x+\frac{e}{k_{0}p_{-}}\int^{\varphi}d\varphi'\left[p_{-}A(\varphi')+\frac{e}{2}A^{2}(\varphi')\right].\label{eq:Sminus}
\end{equation}
The positron solution is obtained by replacing $p_{-}\rightarrow-p_{+}$
and $u\rightarrow v$ where $v$ is the free Dirac positron bispinor.
With this in mind, it is unnecessary to go through the whole derivation
as it is the same as the one for radiation emission which we carried
out in \cite{PhysRevD.100.116001} as we may just replace $p_f \rightarrow p_-$, $p_i\rightarrow-p_+$, $k\rightarrow-k$ and $\epsilon^{*}\rightarrow\epsilon$,
and replace the phase space factors $d^{3}\boldsymbol{k}d^{3}\boldsymbol{p}_{f}\rightarrow d^{3}\boldsymbol{p}_{-}d^{3}\boldsymbol{p}_{+}$.
We consider a vector potential of the form 

\begin{equation}
A^{\mu}=\sum_{j=1}^{2}a_{j}^{\mu}f_{j}(\varphi),\label{eq:potential}
\end{equation}
where the conditions $a_{1}a_{2}=0$ and $a_{j}k_{0}=0$ are satisfied.
In this way we obtain

\begin{align}
dP & =\frac{e^{2}}{4\omega(k_{0}p_{-})(k_{0}p_{+})}\nonumber \\
 & \times\left|\bar{u}\left(A_{0}\slashed{\epsilon}+\sum_{j=1}^{2}A_{1,j}B_{j}+A_{2,j}C_{j}\right)v\right|^{2}d^{3}\boldsymbol{p}_{-},\label{eq:probvolkov}
\end{align}
where

\begin{equation}
B_{j}=\slashed{\epsilon}\frac{e\slashed{k}_{0}\slashed{a}_{j}}{2k_{0}p_{+}}-\frac{e\slashed{a}_{j}\slashed{k}_{0}}{2k_{0}p_{-}}\slashed{\epsilon},\label{eq:Bj}
\end{equation}

\begin{align}
C_{j} & =\frac{e^{2}a_{j}^{2}}{2\left(k_{0}p_{-}\right)\left(k_{0}p_{+}\right)}\left(\epsilon k_{0}\right)\slashed{k}_{0},\label{eq:Cj}
\end{align}

\begin{flalign}
 & A_{n,j}(s,\alpha_{j},\beta_{j})\nonumber \\
 & =\frac{1}{2\pi}\int_{-\infty}^{\infty}d\varphi f_{j}^{n}(\varphi)e^{i\left(s\varphi+\sum_{j=1}^{2}\left[\alpha_{j}F_{j}(\varphi)+\beta_{j}G_{j}(\varphi)\right]\right)},
\end{flalign}

when $n\neq 0$ and with $s=p_{-}k/p_+ k_0$,
\begin{equation}
\alpha_{j}=e\left\{ \frac{p_{+}a_{j}}{k_{0}p_{+}}-\frac{p_{-}a_{j}}{k_{0}p_{-}}\right\} ,
\end{equation}

\begin{equation}
\beta_{j}=-\frac{e^{2}a_{j}^{2}}{2}\left\{ \frac{1}{k_{0}p_{-}}+\frac{1}{k_{0}p_{+}}\right\} ,
\end{equation}

\begin{flalign}
A_{0}(s,\alpha_{j},\beta_{j}) & =-\frac{1}{s}\sum_{j=1}^{2}\left[\alpha_{j}A_{1,j}+\beta_{j}A_{2,j}\right],\label{eq:A0-1}
\end{flalign}

\begin{equation}
F_{j}(\varphi)=\int_{0}^{\varphi}f_{j}(\varphi')d\varphi',\label{eq:F}
\end{equation}

\begin{equation}
G_{j}(\varphi)=\int_{0}^{\varphi}f_{j}^{2}(\varphi')d\varphi'.\label{eq:G}
\end{equation}
\begin{figure}[t]
\includegraphics[width=1\columnwidth]{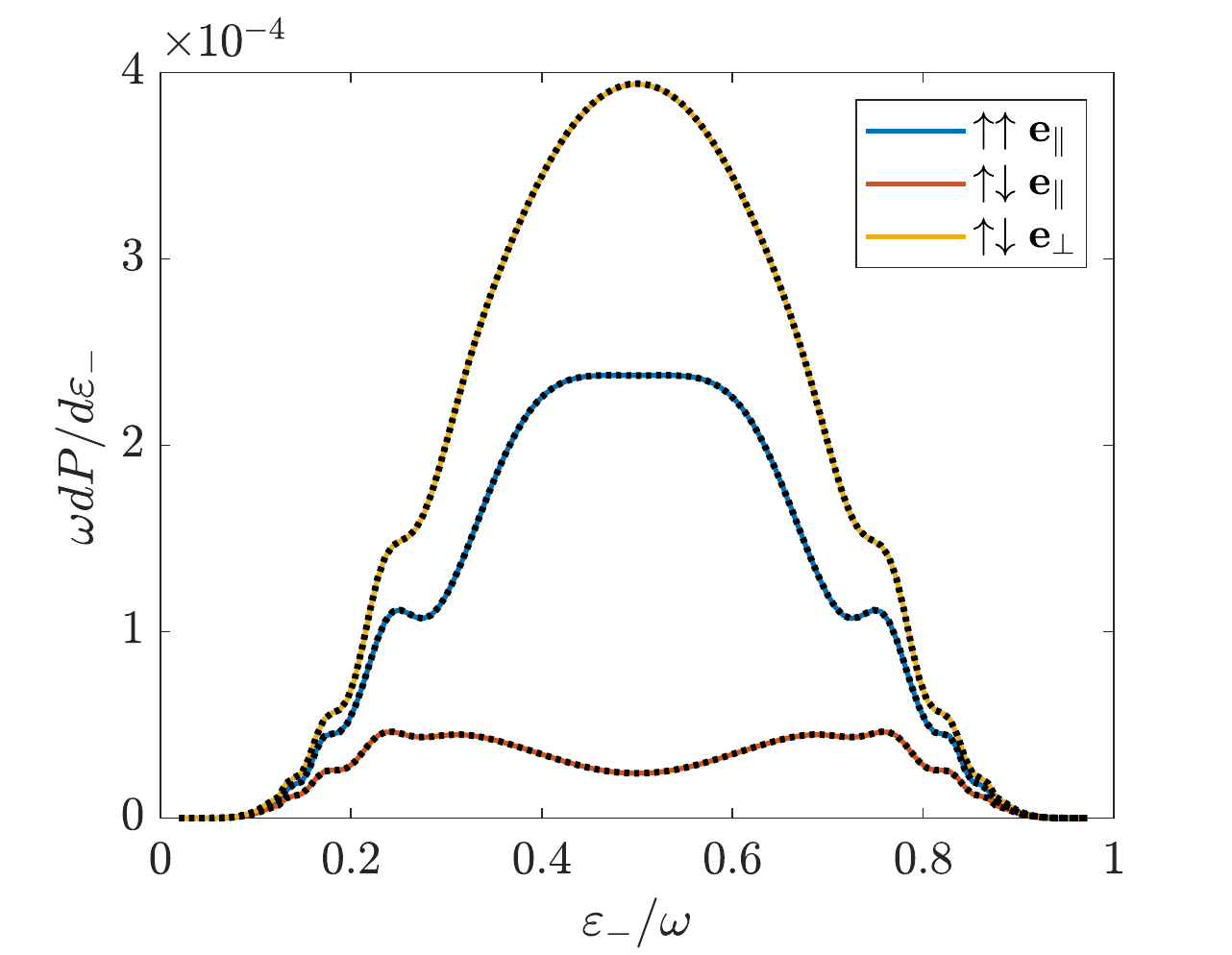}\caption{Here we plot the case of $\xi=1$, $\kappa=1$ and $N=5$ for the
laser pulse described in the text. The fully colored lines is the
result using the semiclassical approach, while the black dots on top
is the same result using the Volkov state. The arrows denote the spin
state of the produced electron and positron respectively, with the
up-arrow denoting spin along the quantization axis $(z)$. The label
$\boldsymbol{e}_{\parallel}$ and $\boldsymbol{e}_{\perp}$ denotes,
respectively, that the incoming photon has polarization parallel or
perpendicular to the polarization of the laser pulse. The remaining
5 possible combinations of spins and polarization are not plotted,
as they coincide with the already plotted curves, however in all cases,
there is agreement.\label{fig:fig1}}
\end{figure}

\section{Discussion}
\begin{figure}[t]
\includegraphics[width=1\columnwidth]{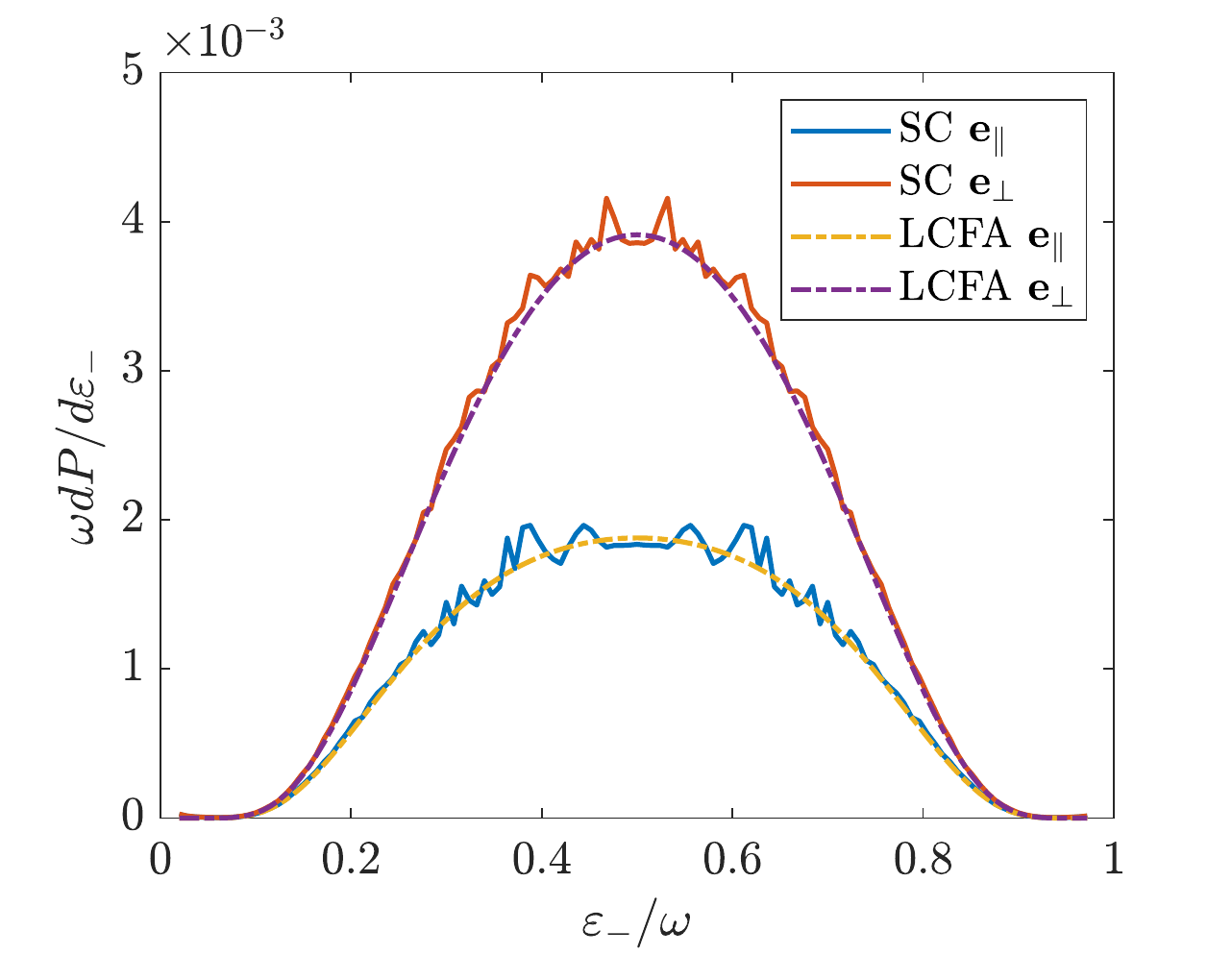}

\caption{Here we plot the case of $\xi=5$, $\kappa=1$ and $N=5$ which has
been summed over the spins, but showing the dependence of the probability
depending on the relative photon polarizations. We compare the full
result with that of the LCFA.\label{fig:fig2}}
\end{figure}
We have now shown how one may calculate the pair production probabilities
using the semiclassical approach, and the approach using the Volkov
state. We will compare the two approaches using an example of a linearly
polarized laser pulse given by $a_{1}^{\mu}=\{0,a_{x},0,0\}$ and
$a_{2}^{\mu}=\{0,0,a_y,0\}$, $k^\mu_{0}=\{\omega_{0},0,0,-\omega_{0}\}$
and $k$ is in the opposite direction of $k_{0}$ and the spin quantization
axis is along the momentum of $k$. We then choose as a model of our
pulse

\begin{equation}
f_{1}(\varphi)=d(\varphi)\text{cos}(\varphi),
\end{equation}

\begin{equation}
f_{2}(\varphi)=0,
\end{equation}

\begin{align}
d(\varphi) & =\begin{cases}
\text{sin}^{4}\left(\frac{\varphi}{2N}\right), & 0<\varphi<2\pi N\\
0. & \text{otherwise}
\end{cases}
\end{align}
We define the invariant quantities $\xi$ and $\kappa$ in terms of
their peak value which leads to the values

\begin{equation}
\xi=\frac{ea_{x}}{m},\label{eq:xsi}
\end{equation}

\begin{equation}
\kappa=\frac{e\sqrt{\left|(F^{\mu\nu}k_{\nu})^{2}\right|}}{m^{3}}=\frac{2\omega\omega_{0}ea_{x}}{m^{3}}\label{eq:kappa}
\end{equation}
where $F^{\mu\nu}$ is the peak value of the electromagnetic field
tensor of the laser pulse. The parameter $\xi$ controls if the process
involves single photons from the external field $(\xi\ll1)$ or many
photons, $(\xi\gg1)$. The $\kappa$ parameter measures the field
experienced by one of the produced particles (if all the energy went
to this particle), relative to the Schwinger critical field, in its
rest frame. When $\kappa$ is small, the pair production process is
heavily suppressed by an exponential ``tunneling'' factor $e^{-8/3\kappa}$, within the local constant field approximation (LCFA),
see e.g. \cite{Ritus}. It is proposed, in the context of the LUXE
experiment, to see this behavior \cite{PhysRevD.99.036008,LUXE}.
It is kept in mind that the experimental setup would involve a target
to produce gamma rays from an electron beam via Bremsstrahlung, which would then collide
with the laser pulse. We consider the case where the gamma ray photon
has the same energy as the initial electron, which is reasonable as
the largest contribution comes from these photons due to the mentioned
tunneling suppression factor. In both of the planned experiments at
SLAC and DESY the pulse duration of around $30$ fs at full width
half maximum of the intensity corresponds to roughly $N=43$ for our
choice of the pulse shape, and therefore we will use this for those
cases along with $\omega_{0}=1.55$ eV. In Fig. \eqref{fig:fig1}
we show an example of $\xi=1$, $\kappa=1$ and $N=5$ where we have
plotted the result from the semiclassical approach along with that
from the Volkov state, and see that the results are indistinguishable.
We have also checked for other values of these parameters, and also for the situtation where the laser beam is not counterpropagating 
with the incoming photon, and in every case there is as good agreement as seen in Fig. \eqref{fig:fig1}.
It has also been checked that the mentioned additional integral which
arises for pair production, but not in radiation emission, plays a
role for the result, and therefore that it has been handled correctly.
In the LUXE experiment it is planned that the first stage of the experiment
is done at $\xi=2$ using a 30 TW laser \cite{Ritus}. In the SLAC
experiment, a 17 TW laser is available which it is envisaged to focus
down to the diffraction limit yielding $\xi=7.3$.
The difference here is therefore that the first stage of the LUXE
experiment is set more conservatively in terms of focusing the laser
pulse. Both experiments plan on achieving $\xi>5$ and therefore in
Fig. \eqref{fig:fig2} we verify that when $\xi=5$ and $\kappa=1$,
one may to good accuracy use the LCFA,
which means using the formula for pair production in a constant crossed
field, which can be found in e.g. \cite{Baier1998}, at each instant of the laser pulse. However for the case of
a potential first stage of these experiments, 
\begin{figure}[t]
\includegraphics[width=1\columnwidth]{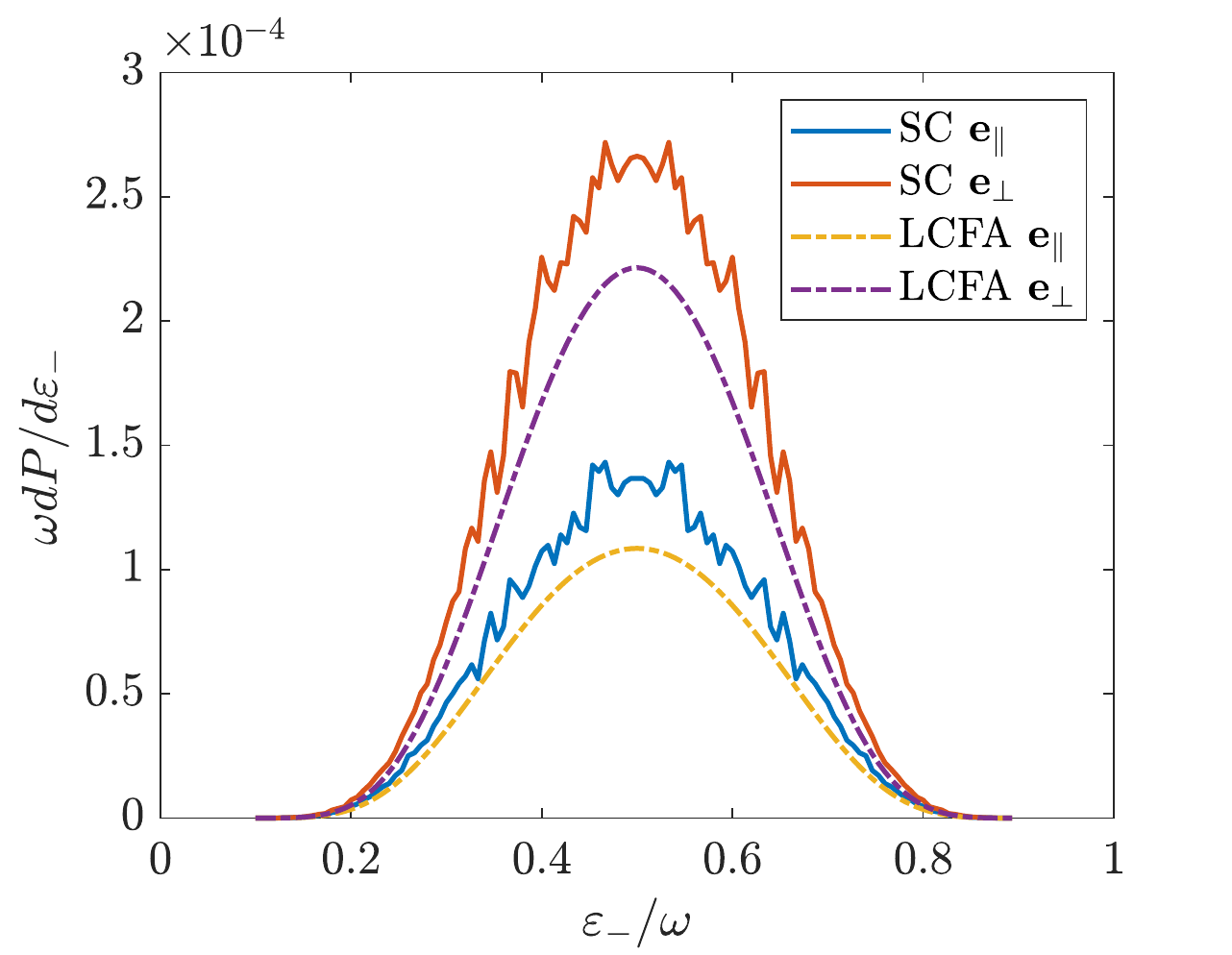}

\caption{Here we show the case of $\xi=2$, $\kappa=0.4$ and $N=43$ which
arises for a photon energy of $17.5$ GeV. This result therefore applies
to the first stage of the proposed LUXE experiment \cite{LUXE}.\label{fig:fig3}}
\end{figure}
where the fully focused pulse may not yet be achieved, we will see
deviations from the LCFA result as $\xi$ goes closer to $1$. As
pointed out by Ritus in \cite{Ritus} in section 13, for the case
of a monochromatic wave, if $\xi$ is not large, or if $\kappa\ll\xi/\sqrt{1+\xi^{2}}$
deviations from the LCFA arise. Most importantly, the overall pair
production rate starts to deviate from the LCFA result. This can be
seen in Fig. \eqref{fig:fig3} and \eqref{fig:fig4}. In general,
the pair production probability is larger than predicted by the LCFA
when $\xi$ approaches $1$ from above. For the case shown in Fig.
\eqref{fig:fig4} the polarization averaged total probability in the
exact case is 27\% larger than that using the LCFA and for the case
in Fig. \eqref{fig:fig3} it is 23\%. An interesting prediction seen
in \cite{Ritus} is that the pair production probability depends on
the relative polarization between the laser pulse and the incoming
photon. In particular it is shown that for the monochromatic wave
and for $\kappa\ll1$ the pair production rate for different relative
polarizations obey $W_{\perp}=2W_{\parallel}$ and $W_{\perp}=3/2W_{\parallel}$
when $\kappa\gg1$. This prediction has not been experimentally verified.
As shown in \cite{Baier1998} the strong field can be seen as a dispersive
medium, where the pair production corresponds to the imaginary part
of the refractive index of this medium, the real part of which can
be obtained by the Kramers-Kronig relations, i.e. the process of vacuum
birefringence in QED. This process has been extensively studied \cite{Heinzl2006a,PhysRevLett.97.083603,1751-8121-40-5-F01,PhysRevD.92.071301,PhysRevD.78.032006,BAKALOV1994180,PhysRevA.94.062102,PhysRevD.88.053009,PhysRevLett.119.250403,Denisov2017,PhysRevD.98.056010,PhysRevA.97.033803},
but not yet experimentally observed. Therefore the clear demonstration
of the pair production rate's dependence on the relative polarization
is an indirect detection of the vacuum birefringence of QED. For the
cases shown in Fig. \eqref{fig:fig3} and \eqref{fig:fig4} we have
that $W_{\perp}=2.05W_{\parallel}$ and $W_{\perp}=2.04W_{\parallel}$,
respectively. However for this measurement one would need polarized
gamma rays, which can be obtained using either Compton back scattering
on a small fraction of the laser pulse, or produced using a crystal
target and the process of coherent bremsstrahlung, see e.g. \cite{Ter-Mikaelian1972}.
A calculation of this is, however, beyond the scope of the current
paper.
\begin{figure}[t]
\includegraphics[width=1\columnwidth]{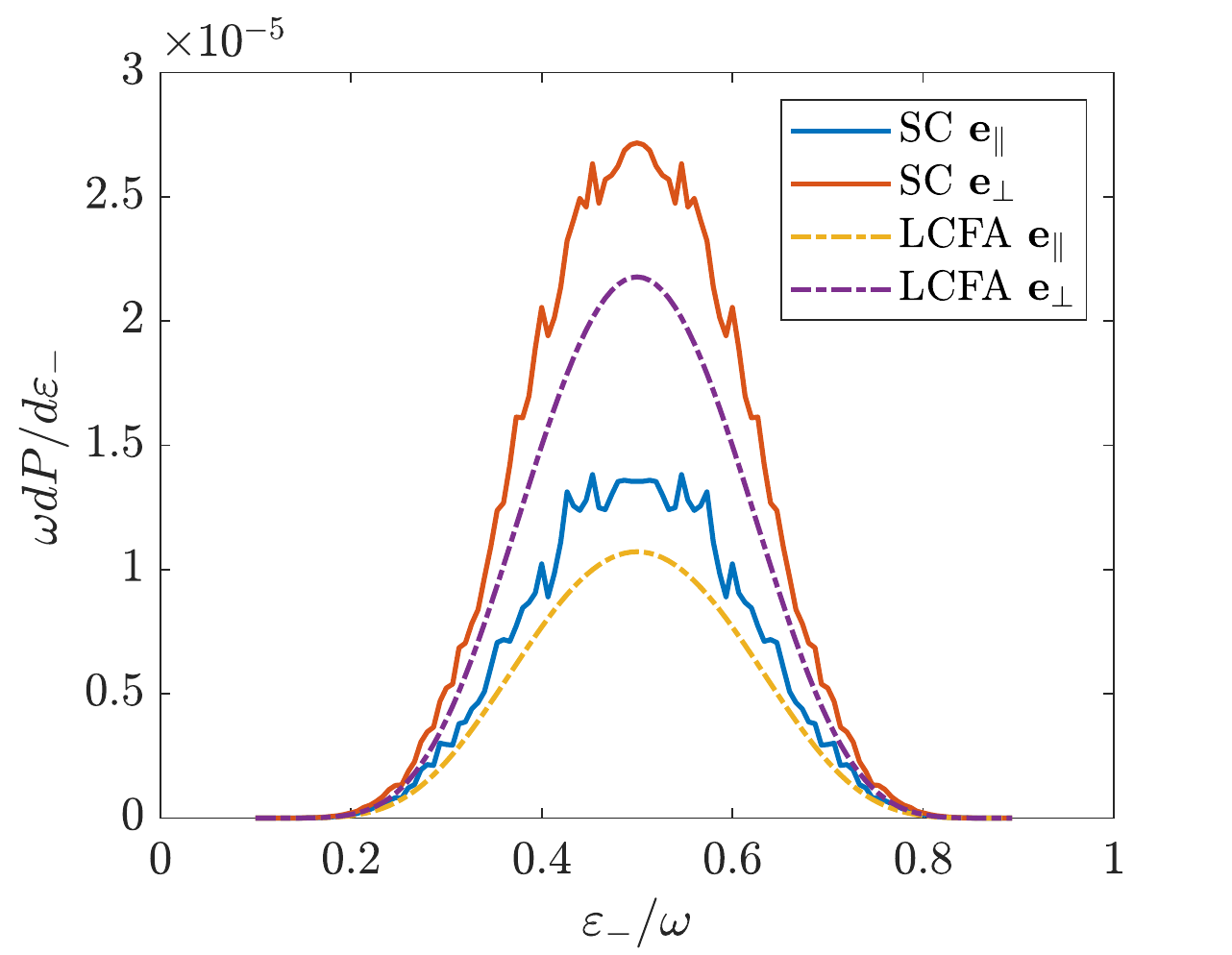}

\caption{Here we show the case of $\xi=2$, $\kappa=0.3$ and $N=43$ which
arises for a photon energy of $13.0$ GeV, which could be observed
in a potential first stage of the SLAC E320 experiment, before the
strongest focusing is achieved.\label{fig:fig4}}
\end{figure}

\section{Conclusion}

We have shown how the semiclassical approach of Baier, Katkov and
Strakhovenko may be recast in a form suitable for numerical implementation,
allowing one to calculate the pair production probability in an arbitrary
external field and for any photon polarization and electron-positron spins.
We compared the results for a case where an exact solution is known,
namely the Volkov state describing an electron-positron in a laser wave. In this case the results are indistinguishable. We investigated the size of the deviations from the local constant
field approximation for experiments planned in the near future, when
$\xi$ is not large. We saw that when $\xi=2$ deviations of around
25\% in the overall rate should be expected.  Finally the presented numerical approach 
allows to study polarization effects in pair production even in complicated fields and we used this to see that the probability
of pair production in the two states of polarization, parallel and
orthogonal to the linearly polarized laser pulse, still yields a factor
of roughly $2$, even though $\kappa$ is not negligible and that we are dealing
with a laser pulse rather than a monochromatic wave.

\section{Acknowledgements}

The author would like to thank Antonino Di Piazza for reading the manuscript and providing useful comments. T. N. Wistisen was supported by the Alexander von Humboldt-Stiftung for this work.

\bibliography{biblio}

\begin{thebibliography}{67}%
\makeatletter
\providecommand \@ifxundefined [1]{%
 \@ifx{#1\undefined}
}%
\providecommand \@ifnum [1]{%
 \ifnum #1\expandafter \@firstoftwo
 \else \expandafter \@secondoftwo
 \fi
}%
\providecommand \@ifx [1]{%
 \ifx #1\expandafter \@firstoftwo
 \else \expandafter \@secondoftwo
 \fi
}%
\providecommand \natexlab [1]{#1}%
\providecommand \enquote  [1]{``#1''}%
\providecommand \bibnamefont  [1]{#1}%
\providecommand \bibfnamefont [1]{#1}%
\providecommand \citenamefont [1]{#1}%
\providecommand \href@noop [0]{\@secondoftwo}%
\providecommand \href [0]{\begingroup \@sanitize@url \@href}%
\providecommand \@href[1]{\@@startlink{#1}\@@href}%
\providecommand \@@href[1]{\endgroup#1\@@endlink}%
\providecommand \@sanitize@url [0]{\catcode `\\12\catcode `\$12\catcode
  `\&12\catcode `\#12\catcode `\^12\catcode `\_12\catcode `\%12\relax}%
\providecommand \@@startlink[1]{}%
\providecommand \@@endlink[0]{}%
\providecommand \url  [0]{\begingroup\@sanitize@url \@url }%
\providecommand \@url [1]{\endgroup\@href {#1}{\urlprefix }}%
\providecommand \urlprefix  [0]{URL }%
\providecommand \Eprint [0]{\href }%
\providecommand \doibase [0]{http://dx.doi.org/}%
\providecommand \selectlanguage [0]{\@gobble}%
\providecommand \bibinfo  [0]{\@secondoftwo}%
\providecommand \bibfield  [0]{\@secondoftwo}%
\providecommand \translation [1]{[#1]}%
\providecommand \BibitemOpen [0]{}%
\providecommand \bibitemStop [0]{}%
\providecommand \bibitemNoStop [0]{.\EOS\space}%
\providecommand \EOS [0]{\spacefactor3000\relax}%
\providecommand \BibitemShut  [1]{\csname bibitem#1\endcsname}%
\let\auto@bib@innerbib\@empty
\bibitem [{\citenamefont {Bula}\ \emph {et~al.}(1996)\citenamefont {Bula},
  \citenamefont {McDonald}, \citenamefont {Prebys}, \citenamefont {Bamber},
  \citenamefont {Boege}, \citenamefont {Kotseroglou}, \citenamefont
  {Melissinos}, \citenamefont {Meyerhofer}, \citenamefont {Ragg}, \citenamefont
  {Burke} \emph {et~al.}}]{bula1996observation}%
  \BibitemOpen
  \bibfield  {author} {\bibinfo {author} {\bibfnamefont {C.}~\bibnamefont
  {Bula}}, \bibinfo {author} {\bibfnamefont {K.T.}\ \bibnamefont {McDonald}},
  \bibinfo {author} {\bibfnamefont {E.J.}\ \bibnamefont {Prebys}}, \bibinfo
  {author} {\bibfnamefont {C.}~\bibnamefont {Bamber}}, \bibinfo {author}
  {\bibfnamefont {S.}~\bibnamefont {Boege}}, \bibinfo {author} {\bibfnamefont
  {T.}~\bibnamefont {Kotseroglou}}, \bibinfo {author} {\bibfnamefont {A.C.}\
  \bibnamefont {Melissinos}}, \bibinfo {author} {\bibfnamefont {D.D.}\
  \bibnamefont {Meyerhofer}}, \bibinfo {author} {\bibfnamefont
  {W.}~\bibnamefont {Ragg}}, \bibinfo {author} {\bibfnamefont {D.L.}\
  \bibnamefont {Burke}},  \emph {et~al.},\ }\bibfield  {title} {\enquote
  {\bibinfo {title} {{Observation of nonlinear effects in Compton
  scattering}},}\ }\href@noop {} {\bibfield  {journal} {\bibinfo  {journal}
  {Phys. Rev. Lett.}\ }\textbf {\bibinfo {volume} {76}},\ \bibinfo {pages}
  {3116} (\bibinfo {year} {1996})}\BibitemShut {NoStop}%
\bibitem [{\citenamefont {Bak}\ \emph {et~al.}(1985)\citenamefont {Bak},
  \citenamefont {Ellison}, \citenamefont {Marsh}, \citenamefont {Meyer},
  \citenamefont {Pedersen}, \citenamefont {Petersen}, \citenamefont
  {Uggerh{\o}j}, \citenamefont {{\O}stergaard}, \citenamefont {M{\o}ller},
  \citenamefont {S{\o}rensen},\ and\ \citenamefont {Suffert}}]{BAK1985491}%
  \BibitemOpen
  \bibfield  {author} {\bibinfo {author} {\bibfnamefont {J.}~\bibnamefont
  {Bak}}, \bibinfo {author} {\bibfnamefont {J.A.}\ \bibnamefont {Ellison}},
  \bibinfo {author} {\bibfnamefont {B.}~\bibnamefont {Marsh}}, \bibinfo
  {author} {\bibfnamefont {F.E.}\ \bibnamefont {Meyer}}, \bibinfo {author}
  {\bibfnamefont {O.}~\bibnamefont {Pedersen}}, \bibinfo {author}
  {\bibfnamefont {J.B.B.}\ \bibnamefont {Petersen}}, \bibinfo {author}
  {\bibfnamefont {E.}~\bibnamefont {Uggerh{\o}j}}, \bibinfo {author}
  {\bibfnamefont {K.}~\bibnamefont {{\O}stergaard}}, \bibinfo {author}
  {\bibfnamefont {S.P.}\ \bibnamefont {M{\o}ller}}, \bibinfo {author}
  {\bibfnamefont {A.H.}\ \bibnamefont {S{\o}rensen}}, \ and\ \bibinfo {author}
  {\bibfnamefont {M.}~\bibnamefont {Suffert}},\ }\bibfield  {title} {\enquote
  {\bibinfo {title} {{Channeling radiation from 2-55 GeV/c electrons and
  positrons: (I). Planar case}},}\ }\href {\doibase
  https://doi.org/10.1016/0550-3213(85)90230-5} {\bibfield  {journal} {\bibinfo
   {journal} {Nucl. Phys. B.}\ }\textbf {\bibinfo {volume} {254}},\ \bibinfo
  {pages} {491 -- 527} (\bibinfo {year} {1985})}\BibitemShut {NoStop}%
\bibitem [{\citenamefont {Bak}\ \emph {et~al.}(1988)\citenamefont {Bak},
  \citenamefont {Ellison}, \citenamefont {Marsh}, \citenamefont {Meyer},
  \citenamefont {Pedersen}, \citenamefont {Petersen}, \citenamefont
  {Uggerh{\o}j}, \citenamefont {M{\o}ller}, \citenamefont {S{\o}rensen},\ and\
  \citenamefont {Suffert}}]{BAK1988525}%
  \BibitemOpen
  \bibfield  {author} {\bibinfo {author} {\bibfnamefont {J.F.}\ \bibnamefont
  {Bak}}, \bibinfo {author} {\bibfnamefont {J.A.}\ \bibnamefont {Ellison}},
  \bibinfo {author} {\bibfnamefont {B.}~\bibnamefont {Marsh}}, \bibinfo
  {author} {\bibfnamefont {F.E.}\ \bibnamefont {Meyer}}, \bibinfo {author}
  {\bibfnamefont {O.}~\bibnamefont {Pedersen}}, \bibinfo {author}
  {\bibfnamefont {J.B.B.}\ \bibnamefont {Petersen}}, \bibinfo {author}
  {\bibfnamefont {E.}~\bibnamefont {Uggerh{\o}j}}, \bibinfo {author}
  {\bibfnamefont {S.P.}\ \bibnamefont {M{\o}ller}}, \bibinfo {author}
  {\bibfnamefont {H.}~\bibnamefont {S{\o}rensen}}, \ and\ \bibinfo {author}
  {\bibfnamefont {M.}~\bibnamefont {Suffert}},\ }\bibfield  {title} {\enquote
  {\bibinfo {title} {{Channeling radiation from 2 to 20 GeV/c electrons and
  positrons (II).: Axial case}},}\ }\href {\doibase
  https://doi.org/10.1016/0550-3213(88)90187-3} {\bibfield  {journal} {\bibinfo
   {journal} {Nucl. Phys. B.}\ }\textbf {\bibinfo {volume} {302}},\ \bibinfo
  {pages} {525 -- 558} (\bibinfo {year} {1988})}\BibitemShut {NoStop}%
\bibitem [{\citenamefont {Swent}\ \emph {et~al.}(1979)\citenamefont {Swent},
  \citenamefont {Pantell}, \citenamefont {Alguard}, \citenamefont {Berman},
  \citenamefont {Bloom},\ and\ \citenamefont {Datz}}]{PhysRevLett.43.1723}%
  \BibitemOpen
  \bibfield  {author} {\bibinfo {author} {\bibfnamefont {R.~L.}\ \bibnamefont
  {Swent}}, \bibinfo {author} {\bibfnamefont {R.~H.}\ \bibnamefont {Pantell}},
  \bibinfo {author} {\bibfnamefont {M.~J.}\ \bibnamefont {Alguard}}, \bibinfo
  {author} {\bibfnamefont {B.~L.}\ \bibnamefont {Berman}}, \bibinfo {author}
  {\bibfnamefont {S.~D.}\ \bibnamefont {Bloom}}, \ and\ \bibinfo {author}
  {\bibfnamefont {S.}~\bibnamefont {Datz}},\ }\bibfield  {title} {\enquote
  {\bibinfo {title} {Observation of channeling radiation from relativistic
  electrons},}\ }\href {\doibase 10.1103/PhysRevLett.43.1723} {\bibfield
  {journal} {\bibinfo  {journal} {Phys. Rev. Lett.}\ }\textbf {\bibinfo
  {volume} {43}},\ \bibinfo {pages} {1723--1726} (\bibinfo {year}
  {1979})}\BibitemShut {NoStop}%
\bibitem [{\citenamefont {Andersen}\ \emph {et~al.}(1981)\citenamefont
  {Andersen}, \citenamefont {Eriksen},\ and\ \citenamefont
  {Laegsgaard}}]{1402-4896-24-3-015}%
  \BibitemOpen
  \bibfield  {author} {\bibinfo {author} {\bibfnamefont {J.U.}\ \bibnamefont
  {Andersen}}, \bibinfo {author} {\bibfnamefont {K.R.}\ \bibnamefont
  {Eriksen}}, \ and\ \bibinfo {author} {\bibfnamefont {E.}~\bibnamefont
  {Laegsgaard}},\ }\bibfield  {title} {\enquote {\bibinfo {title}
  {{Planar-Channeling Radiation and Coherent Bremsstrahlung for MeV
  Electrons}},}\ }\href {http://stacks.iop.org/1402-4896/24/i=3/a=015}
  {\bibfield  {journal} {\bibinfo  {journal} {Phys. Scr.}\ }\textbf {\bibinfo
  {volume} {24}},\ \bibinfo {pages} {588} (\bibinfo {year} {1981})}\BibitemShut
  {NoStop}%
\bibitem [{\citenamefont {Andersen}\ \emph {et~al.}(1982)\citenamefont
  {Andersen}, \citenamefont {Bonderup}, \citenamefont {Laegsgaard},
  \citenamefont {Marsh},\ and\ \citenamefont {S{\o}rensen}}]{ANDERSEN1982209}%
  \BibitemOpen
  \bibfield  {author} {\bibinfo {author} {\bibfnamefont {J.U.}\ \bibnamefont
  {Andersen}}, \bibinfo {author} {\bibfnamefont {E.}~\bibnamefont {Bonderup}},
  \bibinfo {author} {\bibfnamefont {E.}~\bibnamefont {Laegsgaard}}, \bibinfo
  {author} {\bibfnamefont {B.B.}\ \bibnamefont {Marsh}}, \ and\ \bibinfo
  {author} {\bibfnamefont {A.H.}\ \bibnamefont {S{\o}rensen}},\ }\bibfield
  {title} {\enquote {\bibinfo {title} {{Axial channeling radiation from MeV
  electrons}},}\ }\href {\doibase https://doi.org/10.1016/0029-554X(82)90517-1}
  {\bibfield  {journal} {\bibinfo  {journal} {Nucl. Instrum. Methods Phys.
  Res.}\ }\textbf {\bibinfo {volume} {194}},\ \bibinfo {pages} {209 -- 224}
  (\bibinfo {year} {1982})}\BibitemShut {NoStop}%
\bibitem [{\citenamefont {Klein}\ \emph {et~al.}(1985)\citenamefont {Klein},
  \citenamefont {Kephart}, \citenamefont {Pantell}, \citenamefont {Park},
  \citenamefont {Berman}, \citenamefont {Swent}, \citenamefont {Datz},\ and\
  \citenamefont {Fearick}}]{PhysRevB.31.68}%
  \BibitemOpen
  \bibfield  {author} {\bibinfo {author} {\bibfnamefont {R.~K.}\ \bibnamefont
  {Klein}}, \bibinfo {author} {\bibfnamefont {J.~O.}\ \bibnamefont {Kephart}},
  \bibinfo {author} {\bibfnamefont {R.~H.}\ \bibnamefont {Pantell}}, \bibinfo
  {author} {\bibfnamefont {H.}~\bibnamefont {Park}}, \bibinfo {author}
  {\bibfnamefont {B.~L.}\ \bibnamefont {Berman}}, \bibinfo {author}
  {\bibfnamefont {R.~L.}\ \bibnamefont {Swent}}, \bibinfo {author}
  {\bibfnamefont {S.}~\bibnamefont {Datz}}, \ and\ \bibinfo {author}
  {\bibfnamefont {R.~W.}\ \bibnamefont {Fearick}},\ }\bibfield  {title}
  {\enquote {\bibinfo {title} {Electron channeling radiation from diamond},}\
  }\href {\doibase 10.1103/PhysRevB.31.68} {\bibfield  {journal} {\bibinfo
  {journal} {Phys. Rev. B}\ }\textbf {\bibinfo {volume} {31}},\ \bibinfo
  {pages} {68--92} (\bibinfo {year} {1985})}\BibitemShut {NoStop}%
\bibitem [{\citenamefont {Alguard}\ \emph {et~al.}(1979)\citenamefont
  {Alguard}, \citenamefont {Swent}, \citenamefont {Pantell}, \citenamefont
  {Berman}, \citenamefont {Bloom},\ and\ \citenamefont
  {Datz}}]{PhysRevLett.42.1148}%
  \BibitemOpen
  \bibfield  {author} {\bibinfo {author} {\bibfnamefont {M.~J.}\ \bibnamefont
  {Alguard}}, \bibinfo {author} {\bibfnamefont {R.~L.}\ \bibnamefont {Swent}},
  \bibinfo {author} {\bibfnamefont {R.~H.}\ \bibnamefont {Pantell}}, \bibinfo
  {author} {\bibfnamefont {B.~L.}\ \bibnamefont {Berman}}, \bibinfo {author}
  {\bibfnamefont {S.~D.}\ \bibnamefont {Bloom}}, \ and\ \bibinfo {author}
  {\bibfnamefont {S.}~\bibnamefont {Datz}},\ }\bibfield  {title} {\enquote
  {\bibinfo {title} {Observation of radiation from channeled positrons},}\
  }\href {\doibase 10.1103/PhysRevLett.42.1148} {\bibfield  {journal} {\bibinfo
   {journal} {Phys. Rev. Lett.}\ }\textbf {\bibinfo {volume} {42}},\ \bibinfo
  {pages} {1148--1151} (\bibinfo {year} {1979})}\BibitemShut {NoStop}%
\bibitem [{\citenamefont {Andersen}\ \emph {et~al.}(2012)\citenamefont
  {Andersen}, \citenamefont {Esberg}, \citenamefont {Knudsen}, \citenamefont
  {Thomsen}, \citenamefont {Uggerh\o{}j}, \citenamefont {Sona}, \citenamefont
  {Mangiarotti}, \citenamefont {Ketel}, \citenamefont {Dizdar},\ and\
  \citenamefont {Ballestrero}}]{PhysRevD.86.072001}%
  \BibitemOpen
  \bibfield  {author} {\bibinfo {author} {\bibfnamefont {K.~K.}\ \bibnamefont
  {Andersen}}, \bibinfo {author} {\bibfnamefont {J.}~\bibnamefont {Esberg}},
  \bibinfo {author} {\bibfnamefont {H.}~\bibnamefont {Knudsen}}, \bibinfo
  {author} {\bibfnamefont {H.~D.}\ \bibnamefont {Thomsen}}, \bibinfo {author}
  {\bibfnamefont {U.~I.}\ \bibnamefont {Uggerh\o{}j}}, \bibinfo {author}
  {\bibfnamefont {P.}~\bibnamefont {Sona}}, \bibinfo {author} {\bibfnamefont
  {A.}~\bibnamefont {Mangiarotti}}, \bibinfo {author} {\bibfnamefont {T.~J.}\
  \bibnamefont {Ketel}}, \bibinfo {author} {\bibfnamefont {A.}~\bibnamefont
  {Dizdar}}, \ and\ \bibinfo {author} {\bibfnamefont {S.}~\bibnamefont
  {Ballestrero}} (\bibinfo {collaboration} {CERN NA63}),\ }\bibfield  {title}
  {\enquote {\bibinfo {title} {Experimental investigations of synchrotron
  radiation at the onset of the quantum regime},}\ }\href {\doibase
  10.1103/PhysRevD.86.072001} {\bibfield  {journal} {\bibinfo  {journal} {Phys.
  Rev. D}\ }\textbf {\bibinfo {volume} {86}},\ \bibinfo {pages} {072001}
  (\bibinfo {year} {2012})}\BibitemShut {NoStop}%
\bibitem [{\citenamefont {Uggerh\o{}j}(2005)}]{RevModPhys.77.1131}%
  \BibitemOpen
  \bibfield  {author} {\bibinfo {author} {\bibfnamefont {U.~I.}\ \bibnamefont
  {Uggerh\o{}j}},\ }\bibfield  {title} {\enquote {\bibinfo {title} {The
  interaction of relativistic particles with strong crystalline fields},}\
  }\href {\doibase 10.1103/RevModPhys.77.1131} {\bibfield  {journal} {\bibinfo
  {journal} {Rev. Mod. Phys.}\ }\textbf {\bibinfo {volume} {77}},\ \bibinfo
  {pages} {1131--1171} (\bibinfo {year} {2005})}\BibitemShut {NoStop}%
\bibitem [{\citenamefont {Wistisen}\ \emph {et~al.}(2014)\citenamefont
  {Wistisen}, \citenamefont {Andersen}, \citenamefont {Yilmaz}, \citenamefont
  {Mikkelsen}, \citenamefont {Hansen}, \citenamefont {Uggerh{\o}j},
  \citenamefont {Lauth},\ and\ \citenamefont {Backe}}]{PhysRevLett.112.254801}%
  \BibitemOpen
  \bibfield  {author} {\bibinfo {author} {\bibfnamefont {T.~N.}\ \bibnamefont
  {Wistisen}}, \bibinfo {author} {\bibfnamefont {K.~K.}\ \bibnamefont
  {Andersen}}, \bibinfo {author} {\bibfnamefont {S.}~\bibnamefont {Yilmaz}},
  \bibinfo {author} {\bibfnamefont {R.}~\bibnamefont {Mikkelsen}}, \bibinfo
  {author} {\bibfnamefont {J.~L.}\ \bibnamefont {Hansen}}, \bibinfo {author}
  {\bibfnamefont {U.~I.}\ \bibnamefont {Uggerh{\o}j}}, \bibinfo {author}
  {\bibfnamefont {W.}~\bibnamefont {Lauth}}, \ and\ \bibinfo {author}
  {\bibfnamefont {H.}~\bibnamefont {Backe}},\ }\bibfield  {title} {\enquote
  {\bibinfo {title} {Experimental realization of a new type of crystalline
  undulator},}\ }\href {\doibase 10.1103/PhysRevLett.112.254801} {\bibfield
  {journal} {\bibinfo  {journal} {Phys. Rev. Lett.}\ }\textbf {\bibinfo
  {volume} {112}},\ \bibinfo {pages} {254801} (\bibinfo {year}
  {2014})}\BibitemShut {NoStop}%
\bibitem [{\citenamefont {Di~Piazza}\ \emph {et~al.}(2010)\citenamefont
  {Di~Piazza}, \citenamefont {Hatsagortsyan},\ and\ \citenamefont
  {Keitel}}]{PhysRevLett.105.220403}%
  \BibitemOpen
  \bibfield  {author} {\bibinfo {author} {\bibfnamefont {A.}~\bibnamefont
  {Di~Piazza}}, \bibinfo {author} {\bibfnamefont {K.~Z.}\ \bibnamefont
  {Hatsagortsyan}}, \ and\ \bibinfo {author} {\bibfnamefont {C.~H.}\
  \bibnamefont {Keitel}},\ }\bibfield  {title} {\enquote {\bibinfo {title}
  {{Quantum Radiation Reaction Effects in Multiphoton Compton Scattering}},}\
  }\href {\doibase 10.1103/PhysRevLett.105.220403} {\bibfield  {journal}
  {\bibinfo  {journal} {Phys. Rev. Lett.}\ }\textbf {\bibinfo {volume} {105}},\
  \bibinfo {pages} {220403} (\bibinfo {year} {2010})}\BibitemShut {NoStop}%
\bibitem [{\citenamefont {Neitz}\ and\ \citenamefont
  {Di~Piazza}(2013)}]{PhysRevLett.111.054802}%
  \BibitemOpen
  \bibfield  {author} {\bibinfo {author} {\bibfnamefont {N.}~\bibnamefont
  {Neitz}}\ and\ \bibinfo {author} {\bibfnamefont {A.}~\bibnamefont
  {Di~Piazza}},\ }\bibfield  {title} {\enquote {\bibinfo {title} {Stochasticity
  effects in quantum radiation reaction},}\ }\href {\doibase
  10.1103/PhysRevLett.111.054802} {\bibfield  {journal} {\bibinfo  {journal}
  {Phys. Rev. Lett.}\ }\textbf {\bibinfo {volume} {111}},\ \bibinfo {pages}
  {054802} (\bibinfo {year} {2013})}\BibitemShut {NoStop}%
\bibitem [{\citenamefont {Blackburn}\ \emph {et~al.}(2014)\citenamefont
  {Blackburn}, \citenamefont {Ridgers}, \citenamefont {Kirk},\ and\
  \citenamefont {Bell}}]{PhysRevLett.112.015001}%
  \BibitemOpen
  \bibfield  {author} {\bibinfo {author} {\bibfnamefont {T.~G.}\ \bibnamefont
  {Blackburn}}, \bibinfo {author} {\bibfnamefont {C.~P.}\ \bibnamefont
  {Ridgers}}, \bibinfo {author} {\bibfnamefont {J.~G.}\ \bibnamefont {Kirk}}, \
  and\ \bibinfo {author} {\bibfnamefont {A.~R.}\ \bibnamefont {Bell}},\
  }\bibfield  {title} {\enquote {\bibinfo {title} {Quantum radiation reaction
  in laser--electron-beam collisions},}\ }\href {\doibase
  10.1103/PhysRevLett.112.015001} {\bibfield  {journal} {\bibinfo  {journal}
  {Phys. Rev. Lett.}\ }\textbf {\bibinfo {volume} {112}},\ \bibinfo {pages}
  {015001} (\bibinfo {year} {2014})}\BibitemShut {NoStop}%
\bibitem [{\citenamefont {Ji}\ \emph {et~al.}(2014)\citenamefont {Ji},
  \citenamefont {Pukhov}, \citenamefont {Kostyukov}, \citenamefont {Shen},\
  and\ \citenamefont {Akli}}]{PhysRevLett.112.145003}%
  \BibitemOpen
  \bibfield  {author} {\bibinfo {author} {\bibfnamefont {L.~L.}\ \bibnamefont
  {Ji}}, \bibinfo {author} {\bibfnamefont {A.}~\bibnamefont {Pukhov}}, \bibinfo
  {author} {\bibfnamefont {I.~Yu.}\ \bibnamefont {Kostyukov}}, \bibinfo
  {author} {\bibfnamefont {B.~F.}\ \bibnamefont {Shen}}, \ and\ \bibinfo
  {author} {\bibfnamefont {K.}~\bibnamefont {Akli}},\ }\bibfield  {title}
  {\enquote {\bibinfo {title} {Radiation-reaction trapping of electrons in
  extreme laser fields},}\ }\href {\doibase 10.1103/PhysRevLett.112.145003}
  {\bibfield  {journal} {\bibinfo  {journal} {Phys. Rev. Lett.}\ }\textbf
  {\bibinfo {volume} {112}},\ \bibinfo {pages} {145003} (\bibinfo {year}
  {2014})}\BibitemShut {NoStop}%
\bibitem [{\citenamefont {Ilderton}\ and\ \citenamefont
  {Torgrimsson}(2013)}]{ILDERTON2013481}%
  \BibitemOpen
  \bibfield  {author} {\bibinfo {author} {\bibfnamefont {A.}~\bibnamefont
  {Ilderton}}\ and\ \bibinfo {author} {\bibfnamefont {G.}~\bibnamefont
  {Torgrimsson}},\ }\bibfield  {title} {\enquote {\bibinfo {title} {{Radiation
  reaction in strong field QED}},}\ }\href {\doibase
  https://doi.org/10.1016/j.physletb.2013.07.045} {\bibfield  {journal}
  {\bibinfo  {journal} {Physics Letters B}\ }\textbf {\bibinfo {volume}
  {725}},\ \bibinfo {pages} {481 -- 486} (\bibinfo {year} {2013})}\BibitemShut
  {NoStop}%
\bibitem [{\citenamefont {Dinu}\ \emph {et~al.}(2016)\citenamefont {Dinu},
  \citenamefont {Harvey}, \citenamefont {Ilderton}, \citenamefont {Marklund},\
  and\ \citenamefont {Torgrimsson}}]{PhysRevLett.116.044801}%
  \BibitemOpen
  \bibfield  {author} {\bibinfo {author} {\bibfnamefont {V.}~\bibnamefont
  {Dinu}}, \bibinfo {author} {\bibfnamefont {C.}~\bibnamefont {Harvey}},
  \bibinfo {author} {\bibfnamefont {A.}~\bibnamefont {Ilderton}}, \bibinfo
  {author} {\bibfnamefont {M.}~\bibnamefont {Marklund}}, \ and\ \bibinfo
  {author} {\bibfnamefont {G.}~\bibnamefont {Torgrimsson}},\ }\bibfield
  {title} {\enquote {\bibinfo {title} {Quantum radiation reaction: From
  interference to incoherence},}\ }\href {\doibase
  10.1103/PhysRevLett.116.044801} {\bibfield  {journal} {\bibinfo  {journal}
  {Phys. Rev. Lett.}\ }\textbf {\bibinfo {volume} {116}},\ \bibinfo {pages}
  {044801} (\bibinfo {year} {2016})}\BibitemShut {NoStop}%
\bibitem [{\citenamefont {Vranic}\ \emph {et~al.}(2014)\citenamefont {Vranic},
  \citenamefont {Martins}, \citenamefont {Vieira}, \citenamefont {Fonseca},\
  and\ \citenamefont {Silva}}]{PhysRevLett.113.134801}%
  \BibitemOpen
  \bibfield  {author} {\bibinfo {author} {\bibfnamefont {M.}~\bibnamefont
  {Vranic}}, \bibinfo {author} {\bibfnamefont {J.~L.}\ \bibnamefont {Martins}},
  \bibinfo {author} {\bibfnamefont {J.}~\bibnamefont {Vieira}}, \bibinfo
  {author} {\bibfnamefont {R.~A.}\ \bibnamefont {Fonseca}}, \ and\ \bibinfo
  {author} {\bibfnamefont {L.~O.}\ \bibnamefont {Silva}},\ }\bibfield  {title}
  {\enquote {\bibinfo {title} {All-optical radiation reaction at
  $1{0}^{21}\text{ }\text{ }\mathrm{W}/{\mathrm{cm}}^{2}$},}\ }\href {\doibase
  10.1103/PhysRevLett.113.134801} {\bibfield  {journal} {\bibinfo  {journal}
  {Phys. Rev. Lett.}\ }\textbf {\bibinfo {volume} {113}},\ \bibinfo {pages}
  {134801} (\bibinfo {year} {2014})}\BibitemShut {NoStop}%
\bibitem [{\citenamefont {Seipt}\ and\ \citenamefont
  {K\"ampfer}(2012)}]{PhysRevD.85.101701}%
  \BibitemOpen
  \bibfield  {author} {\bibinfo {author} {\bibfnamefont {D.}~\bibnamefont
  {Seipt}}\ and\ \bibinfo {author} {\bibfnamefont {B.}~\bibnamefont
  {K\"ampfer}},\ }\bibfield  {title} {\enquote {\bibinfo {title} {{Two-photon
  Compton process in pulsed intense laser fields}},}\ }\href {\doibase
  10.1103/PhysRevD.85.101701} {\bibfield  {journal} {\bibinfo  {journal} {Phys.
  Rev. D}\ }\textbf {\bibinfo {volume} {85}},\ \bibinfo {pages} {101701}
  (\bibinfo {year} {2012})}\BibitemShut {NoStop}%
\bibitem [{\citenamefont {Mackenroth}\ and\ \citenamefont
  {Di~Piazza}(2013)}]{PhysRevLett.110.070402}%
  \BibitemOpen
  \bibfield  {author} {\bibinfo {author} {\bibfnamefont {F.}~\bibnamefont
  {Mackenroth}}\ and\ \bibinfo {author} {\bibfnamefont {A.}~\bibnamefont
  {Di~Piazza}},\ }\bibfield  {title} {\enquote {\bibinfo {title} {{Nonlinear
  Double Compton Scattering in the Ultrarelativistic Quantum Regime}},}\ }\href
  {\doibase 10.1103/PhysRevLett.110.070402} {\bibfield  {journal} {\bibinfo
  {journal} {Phys. Rev. Lett.}\ }\textbf {\bibinfo {volume} {110}},\ \bibinfo
  {pages} {070402} (\bibinfo {year} {2013})}\BibitemShut {NoStop}%
\bibitem [{\citenamefont {Dinu}\ and\ \citenamefont
  {Torgrimsson}(2019{\natexlab{a}})}]{dinu2018single}%
  \BibitemOpen
  \bibfield  {author} {\bibinfo {author} {\bibfnamefont {V.}~\bibnamefont
  {Dinu}}\ and\ \bibinfo {author} {\bibfnamefont {G.}~\bibnamefont
  {Torgrimsson}},\ }\bibfield  {title} {\enquote {\bibinfo {title} {{Single and
  double nonlinear Compton scattering}},}\ }\href {\doibase
  10.1103/PhysRevD.99.096018} {\bibfield  {journal} {\bibinfo  {journal} {Phys.
  Rev. D}\ }\textbf {\bibinfo {volume} {99}},\ \bibinfo {pages} {096018}
  (\bibinfo {year} {2019}{\natexlab{a}})}\BibitemShut {NoStop}%
\bibitem [{\citenamefont {King}(2015)}]{PhysRevA.91.033415}%
  \BibitemOpen
  \bibfield  {author} {\bibinfo {author} {\bibfnamefont {B.}~\bibnamefont
  {King}},\ }\bibfield  {title} {\enquote {\bibinfo {title} {{Double Compton
  scattering in a constant crossed field}},}\ }\href {\doibase
  10.1103/PhysRevA.91.033415} {\bibfield  {journal} {\bibinfo  {journal} {Phys.
  Rev. A}\ }\textbf {\bibinfo {volume} {91}},\ \bibinfo {pages} {033415}
  (\bibinfo {year} {2015})}\BibitemShut {NoStop}%
\bibitem [{\citenamefont {Dinu}\ and\ \citenamefont
  {Torgrimsson}(2019{\natexlab{b}})}]{dinu2019approximating}%
  \BibitemOpen
  \bibfield  {author} {\bibinfo {author} {\bibfnamefont {V.}~\bibnamefont
  {Dinu}}\ and\ \bibinfo {author} {\bibfnamefont {G.}~\bibnamefont
  {Torgrimsson}},\ }\bibfield  {title} {\enquote {\bibinfo {title}
  {{Approximating higher-order nonlinear QED processes with first-order
  building blocks}},}\ }\href@noop {} {\  (\bibinfo {year}
  {2019}{\natexlab{b}})},\ \Eprint {http://arxiv.org/abs/1912.11015}
  {arXiv:1912.11015 [hep-ph]} \BibitemShut {NoStop}%
\bibitem [{\citenamefont {Wistisen}\ \emph {et~al.}(2018)\citenamefont
  {Wistisen}, \citenamefont {Di~Piazza}, \citenamefont {Knudsen},\ and\
  \citenamefont {Uggerh{\o}j}}]{Wistisen2018experimental}%
  \BibitemOpen
  \bibfield  {author} {\bibinfo {author} {\bibfnamefont {T.~N.}\ \bibnamefont
  {Wistisen}}, \bibinfo {author} {\bibfnamefont {A.}~\bibnamefont {Di~Piazza}},
  \bibinfo {author} {\bibfnamefont {H.~V.}\ \bibnamefont {Knudsen}}, \ and\
  \bibinfo {author} {\bibfnamefont {U.~I.}\ \bibnamefont {Uggerh{\o}j}},\
  }\bibfield  {title} {\enquote {\bibinfo {title} {Experimental evidence of
  quantum radiation reaction in aligned crystals},}\ }\href
  {https://doi.org/10.1038/s41467-018-03165-4} {\bibfield  {journal} {\bibinfo
  {journal} {Nat. Commun.}\ }\textbf {\bibinfo {volume} {9}},\ \bibinfo {pages}
  {795} (\bibinfo {year} {2018})}\BibitemShut {NoStop}%
\bibitem [{\citenamefont {Cole}\ \emph {et~al.}(2018)\citenamefont {Cole},
  \citenamefont {Behm}, \citenamefont {Gerstmayr}, \citenamefont {Blackburn},
  \citenamefont {Wood}, \citenamefont {Baird}, \citenamefont {Duff},
  \citenamefont {Harvey}, \citenamefont {Ilderton}, \citenamefont {Joglekar},
  \citenamefont {Krushelnick}, \citenamefont {Kuschel}, \citenamefont
  {Marklund}, \citenamefont {McKenna}, \citenamefont {Murphy}, \citenamefont
  {Poder}, \citenamefont {Ridgers}, \citenamefont {Samarin}, \citenamefont
  {Sarri}, \citenamefont {Symes}, \citenamefont {Thomas}, \citenamefont
  {Warwick}, \citenamefont {Zepf}, \citenamefont {Najmudin},\ and\
  \citenamefont {Mangles}}]{PhysRevX.8.011020}%
  \BibitemOpen
  \bibfield  {author} {\bibinfo {author} {\bibfnamefont {J.~M.}\ \bibnamefont
  {Cole}}, \bibinfo {author} {\bibfnamefont {K.~T.}\ \bibnamefont {Behm}},
  \bibinfo {author} {\bibfnamefont {E.}~\bibnamefont {Gerstmayr}}, \bibinfo
  {author} {\bibfnamefont {T.~G.}\ \bibnamefont {Blackburn}}, \bibinfo {author}
  {\bibfnamefont {J.~C.}\ \bibnamefont {Wood}}, \bibinfo {author}
  {\bibfnamefont {C.~D.}\ \bibnamefont {Baird}}, \bibinfo {author}
  {\bibfnamefont {M.~J.}\ \bibnamefont {Duff}}, \bibinfo {author}
  {\bibfnamefont {C.}~\bibnamefont {Harvey}}, \bibinfo {author} {\bibfnamefont
  {A.}~\bibnamefont {Ilderton}}, \bibinfo {author} {\bibfnamefont {A.~S.}\
  \bibnamefont {Joglekar}}, \bibinfo {author} {\bibfnamefont {K.}~\bibnamefont
  {Krushelnick}}, \bibinfo {author} {\bibfnamefont {S.}~\bibnamefont
  {Kuschel}}, \bibinfo {author} {\bibfnamefont {M.}~\bibnamefont {Marklund}},
  \bibinfo {author} {\bibfnamefont {P.}~\bibnamefont {McKenna}}, \bibinfo
  {author} {\bibfnamefont {C.~D.}\ \bibnamefont {Murphy}}, \bibinfo {author}
  {\bibfnamefont {K.}~\bibnamefont {Poder}}, \bibinfo {author} {\bibfnamefont
  {C.~P.}\ \bibnamefont {Ridgers}}, \bibinfo {author} {\bibfnamefont {G.~M.}\
  \bibnamefont {Samarin}}, \bibinfo {author} {\bibfnamefont {G.}~\bibnamefont
  {Sarri}}, \bibinfo {author} {\bibfnamefont {D.~R.}\ \bibnamefont {Symes}},
  \bibinfo {author} {\bibfnamefont {A.~G.~R.}\ \bibnamefont {Thomas}}, \bibinfo
  {author} {\bibfnamefont {J.}~\bibnamefont {Warwick}}, \bibinfo {author}
  {\bibfnamefont {M.}~\bibnamefont {Zepf}}, \bibinfo {author} {\bibfnamefont
  {Z.}~\bibnamefont {Najmudin}}, \ and\ \bibinfo {author} {\bibfnamefont
  {S.~P.~D.}\ \bibnamefont {Mangles}},\ }\bibfield  {title} {\enquote {\bibinfo
  {title} {Experimental evidence of radiation reaction in the collision of a
  high-intensity laser pulse with a laser-wakefield accelerated electron
  beam},}\ }\href {\doibase 10.1103/PhysRevX.8.011020} {\bibfield  {journal}
  {\bibinfo  {journal} {Phys. Rev. X}\ }\textbf {\bibinfo {volume} {8}},\
  \bibinfo {pages} {011020} (\bibinfo {year} {2018})}\BibitemShut {NoStop}%
\bibitem [{\citenamefont {Poder}\ \emph {et~al.}(2018)\citenamefont {Poder},
  \citenamefont {Tamburini}, \citenamefont {Sarri}, \citenamefont {Di~Piazza},
  \citenamefont {Kuschel}, \citenamefont {Baird}, \citenamefont {Behm},
  \citenamefont {Bohlen}, \citenamefont {Cole}, \citenamefont {Corvan},
  \citenamefont {Duff}, \citenamefont {Gerstmayr}, \citenamefont {Keitel},
  \citenamefont {Krushelnick}, \citenamefont {Mangles}, \citenamefont
  {McKenna}, \citenamefont {Murphy}, \citenamefont {Najmudin}, \citenamefont
  {Ridgers}, \citenamefont {Samarin}, \citenamefont {Symes}, \citenamefont
  {Thomas}, \citenamefont {Warwick},\ and\ \citenamefont
  {Zepf}}]{PhysRevX.8.031004}%
  \BibitemOpen
  \bibfield  {author} {\bibinfo {author} {\bibfnamefont {K.}~\bibnamefont
  {Poder}}, \bibinfo {author} {\bibfnamefont {M.}~\bibnamefont {Tamburini}},
  \bibinfo {author} {\bibfnamefont {G.}~\bibnamefont {Sarri}}, \bibinfo
  {author} {\bibfnamefont {A.}~\bibnamefont {Di~Piazza}}, \bibinfo {author}
  {\bibfnamefont {S.}~\bibnamefont {Kuschel}}, \bibinfo {author} {\bibfnamefont
  {C.~D.}\ \bibnamefont {Baird}}, \bibinfo {author} {\bibfnamefont
  {K.}~\bibnamefont {Behm}}, \bibinfo {author} {\bibfnamefont {S.}~\bibnamefont
  {Bohlen}}, \bibinfo {author} {\bibfnamefont {J.~M.}\ \bibnamefont {Cole}},
  \bibinfo {author} {\bibfnamefont {D.~J.}\ \bibnamefont {Corvan}}, \bibinfo
  {author} {\bibfnamefont {M.}~\bibnamefont {Duff}}, \bibinfo {author}
  {\bibfnamefont {E.}~\bibnamefont {Gerstmayr}}, \bibinfo {author}
  {\bibfnamefont {C.~H.}\ \bibnamefont {Keitel}}, \bibinfo {author}
  {\bibfnamefont {K.}~\bibnamefont {Krushelnick}}, \bibinfo {author}
  {\bibfnamefont {S.~P.~D.}\ \bibnamefont {Mangles}}, \bibinfo {author}
  {\bibfnamefont {P.}~\bibnamefont {McKenna}}, \bibinfo {author} {\bibfnamefont
  {C.~D.}\ \bibnamefont {Murphy}}, \bibinfo {author} {\bibfnamefont
  {Z.}~\bibnamefont {Najmudin}}, \bibinfo {author} {\bibfnamefont {C.~P.}\
  \bibnamefont {Ridgers}}, \bibinfo {author} {\bibfnamefont {G.~M.}\
  \bibnamefont {Samarin}}, \bibinfo {author} {\bibfnamefont {D.~R.}\
  \bibnamefont {Symes}}, \bibinfo {author} {\bibfnamefont {A.~G.~R.}\
  \bibnamefont {Thomas}}, \bibinfo {author} {\bibfnamefont {J.}~\bibnamefont
  {Warwick}}, \ and\ \bibinfo {author} {\bibfnamefont {M.}~\bibnamefont
  {Zepf}},\ }\bibfield  {title} {\enquote {\bibinfo {title} {Experimental
  signatures of the quantum nature of radiation reaction in the field of an
  ultraintense laser},}\ }\href {\doibase 10.1103/PhysRevX.8.031004} {\bibfield
   {journal} {\bibinfo  {journal} {Phys. Rev. X}\ }\textbf {\bibinfo {volume}
  {8}},\ \bibinfo {pages} {031004} (\bibinfo {year} {2018})}\BibitemShut
  {NoStop}%
\bibitem [{\citenamefont {Wistisen}\ \emph {et~al.}(2019)\citenamefont
  {Wistisen}, \citenamefont {Di~Piazza}, \citenamefont {Nielsen}, \citenamefont
  {S\o{}rensen},\ and\ \citenamefont {Uggerh\o{}j}}]{Wistisen2019exp}%
  \BibitemOpen
  \bibfield  {author} {\bibinfo {author} {\bibfnamefont {T.~N.}\ \bibnamefont
  {Wistisen}}, \bibinfo {author} {\bibfnamefont {A.}~\bibnamefont {Di~Piazza}},
  \bibinfo {author} {\bibfnamefont {C.~F.}\ \bibnamefont {Nielsen}}, \bibinfo
  {author} {\bibfnamefont {A.~H.}\ \bibnamefont {S\o{}rensen}}, \ and\ \bibinfo
  {author} {\bibfnamefont {U.~I.}\ \bibnamefont {Uggerh\o{}j}} (\bibinfo
  {collaboration} {CERN NA63}),\ }\bibfield  {title} {\enquote {\bibinfo
  {title} {Quantum radiation reaction in aligned crystals beyond the local
  constant field approximation},}\ }\href {\doibase
  10.1103/PhysRevResearch.1.033014} {\bibfield  {journal} {\bibinfo  {journal}
  {Phys. Rev. Research}\ }\textbf {\bibinfo {volume} {1}},\ \bibinfo {pages}
  {033014} (\bibinfo {year} {2019})}\BibitemShut {NoStop}%
\bibitem [{\citenamefont {Abramowicz}\ \emph {et~al.}(2019)\citenamefont
  {Abramowicz} \emph {et~al.}}]{LUXE}%
  \BibitemOpen
  \bibfield  {author} {\bibinfo {author} {\bibfnamefont {H.}~\bibnamefont
  {Abramowicz}} \emph {et~al.},\ }\bibfield  {title} {\enquote {\bibinfo
  {title} {{Letter of Intent for the LUXE Experiment}},}\ }\href@noop {} {\
  (\bibinfo {year} {2019})},\ \Eprint {http://arxiv.org/abs/1909.00860}
  {arXiv:1909.00860 [physics.ins-det]} \BibitemShut {NoStop}%
\bibitem [{\citenamefont {Meuren}(2019)}]{FACET}%
  \BibitemOpen
  \bibfield  {author} {\bibinfo {author} {\bibfnamefont {S.}~\bibnamefont
  {Meuren}},\ }\bibfield  {title} {\enquote {\bibinfo {title} {{Probing
  Strong-field QED at FACET-II (SLAC E-320)}},}\ }\href
  {https://conf.slac.stanford.edu/facet-2-2019/sites/facet-2-2019.conf.slac.stanford.edu/files/basic-page-docs/sfqed_2019.pdf}
  {\  (\bibinfo {year} {2019})}\BibitemShut {NoStop}%
\bibitem [{\citenamefont {Turcu}\ \emph {et~al.}(2016)\citenamefont {Turcu}
  \emph {et~al.}}]{Turcu2016}%
  \BibitemOpen
  \bibfield  {author} {\bibinfo {author} {\bibfnamefont {I.~C.~E.}\
  \bibnamefont {Turcu}} \emph {et~al.},\ }\bibfield  {title} {\enquote
  {\bibinfo {title} {{High field physics and QED experiments at ELI-NP}},}\
  }\href@noop {} {\bibfield  {journal} {\bibinfo  {journal} {Rom. Rep. Phys.}\
  }\textbf {\bibinfo {volume} {68}},\ \bibinfo {pages} {S145} (\bibinfo {year}
  {2016})}\BibitemShut {NoStop}%
\bibitem [{\citenamefont {Breit}\ and\ \citenamefont
  {Wheeler}(1934)}]{PhysRev.46.1087}%
  \BibitemOpen
  \bibfield  {author} {\bibinfo {author} {\bibfnamefont {G.}~\bibnamefont
  {Breit}}\ and\ \bibinfo {author} {\bibfnamefont {John~A.}\ \bibnamefont
  {Wheeler}},\ }\bibfield  {title} {\enquote {\bibinfo {title} {Collision of
  two light quanta},}\ }\href {\doibase 10.1103/PhysRev.46.1087} {\bibfield
  {journal} {\bibinfo  {journal} {Phys. Rev.}\ }\textbf {\bibinfo {volume}
  {46}},\ \bibinfo {pages} {1087--1091} (\bibinfo {year} {1934})}\BibitemShut
  {NoStop}%
\bibitem [{\citenamefont {Krajewska}\ and\ \citenamefont
  {Kami\ifmmode~\acute{n}\else \'{n}\fi{}ski}(2012)}]{PhysRevA.86.052104}%
  \BibitemOpen
  \bibfield  {author} {\bibinfo {author} {\bibfnamefont {K.}~\bibnamefont
  {Krajewska}}\ and\ \bibinfo {author} {\bibfnamefont {J.~Z.}\ \bibnamefont
  {Kami\ifmmode~\acute{n}\else \'{n}\fi{}ski}},\ }\bibfield  {title} {\enquote
  {\bibinfo {title} {{Breit-Wheeler process in intense short laser pulses}},}\
  }\href {\doibase 10.1103/PhysRevA.86.052104} {\bibfield  {journal} {\bibinfo
  {journal} {Phys. Rev. A}\ }\textbf {\bibinfo {volume} {86}},\ \bibinfo
  {pages} {052104} (\bibinfo {year} {2012})}\BibitemShut {NoStop}%
\bibitem [{\citenamefont {Titov}\ \emph {et~al.}(2013)\citenamefont {Titov},
  \citenamefont {K\"ampfer}, \citenamefont {Takabe},\ and\ \citenamefont
  {Hosaka}}]{PhysRevA.87.042106}%
  \BibitemOpen
  \bibfield  {author} {\bibinfo {author} {\bibfnamefont {A.~I.}\ \bibnamefont
  {Titov}}, \bibinfo {author} {\bibfnamefont {B.}~\bibnamefont {K\"ampfer}},
  \bibinfo {author} {\bibfnamefont {H.}~\bibnamefont {Takabe}}, \ and\ \bibinfo
  {author} {\bibfnamefont {A.}~\bibnamefont {Hosaka}},\ }\bibfield  {title}
  {\enquote {\bibinfo {title} {{Breit-Wheeler process in very short
  electromagnetic pulses}},}\ }\href {\doibase 10.1103/PhysRevA.87.042106}
  {\bibfield  {journal} {\bibinfo  {journal} {Phys. Rev. A}\ }\textbf {\bibinfo
  {volume} {87}},\ \bibinfo {pages} {042106} (\bibinfo {year}
  {2013})}\BibitemShut {NoStop}%
\bibitem [{\citenamefont {Di~Piazza}(2016)}]{PhysRevLett.117.213201}%
  \BibitemOpen
  \bibfield  {author} {\bibinfo {author} {\bibfnamefont {A.}~\bibnamefont
  {Di~Piazza}},\ }\bibfield  {title} {\enquote {\bibinfo {title} {{Nonlinear
  Breit-Wheeler Pair Production in a Tightly Focused Laser Beam}},}\ }\href
  {\doibase 10.1103/PhysRevLett.117.213201} {\bibfield  {journal} {\bibinfo
  {journal} {Phys. Rev. Lett.}\ }\textbf {\bibinfo {volume} {117}},\ \bibinfo
  {pages} {213201} (\bibinfo {year} {2016})}\BibitemShut {NoStop}%
\bibitem [{\citenamefont {Heinzl}\ \emph {et~al.}(2010)\citenamefont {Heinzl},
  \citenamefont {Ilderton},\ and\ \citenamefont {Marklund}}]{Heinzl2010250}%
  \BibitemOpen
  \bibfield  {author} {\bibinfo {author} {\bibfnamefont {T.}~\bibnamefont
  {Heinzl}}, \bibinfo {author} {\bibfnamefont {A.}~\bibnamefont {Ilderton}}, \
  and\ \bibinfo {author} {\bibfnamefont {M.}~\bibnamefont {Marklund}},\
  }\bibfield  {title} {\enquote {\bibinfo {title} {Finite size effects in
  stimulated laser pair production},}\ }\href {\doibase
  http://dx.doi.org/10.1016/j.physletb.2010.07.044} {\bibfield  {journal}
  {\bibinfo  {journal} {Phys. Lett. B}\ }\textbf {\bibinfo {volume} {692}},\
  \bibinfo {pages} {250 -- 256} (\bibinfo {year} {2010})}\BibitemShut {NoStop}%
\bibitem [{\citenamefont {Titov}\ \emph {et~al.}(2012)\citenamefont {Titov},
  \citenamefont {Takabe}, \citenamefont {K\"ampfer},\ and\ \citenamefont
  {Hosaka}}]{PhysRevLett.108.240406}%
  \BibitemOpen
  \bibfield  {author} {\bibinfo {author} {\bibfnamefont {A.~I.}\ \bibnamefont
  {Titov}}, \bibinfo {author} {\bibfnamefont {H.}~\bibnamefont {Takabe}},
  \bibinfo {author} {\bibfnamefont {B.}~\bibnamefont {K\"ampfer}}, \ and\
  \bibinfo {author} {\bibfnamefont {A.}~\bibnamefont {Hosaka}},\ }\bibfield
  {title} {\enquote {\bibinfo {title} {Enhanced subthreshold
  ${e}^{\mathbf{+}}{e}^{\mathbf{\ensuremath{-}}}$ production in short laser
  pulses},}\ }\href {\doibase 10.1103/PhysRevLett.108.240406} {\bibfield
  {journal} {\bibinfo  {journal} {Phys. Rev. Lett.}\ }\textbf {\bibinfo
  {volume} {108}},\ \bibinfo {pages} {240406} (\bibinfo {year}
  {2012})}\BibitemShut {NoStop}%
\bibitem [{\citenamefont {Jansen}\ and\ \citenamefont
  {M\"uller}(2016)}]{PhysRevD.93.053011}%
  \BibitemOpen
  \bibfield  {author} {\bibinfo {author} {\bibfnamefont {M.~J.~A.}\
  \bibnamefont {Jansen}}\ and\ \bibinfo {author} {\bibfnamefont
  {C.}~\bibnamefont {M\"uller}},\ }\bibfield  {title} {\enquote {\bibinfo
  {title} {{Strong-field Breit-Wheeler pair production in short laser pulses:
  Identifying multiphoton interference and carrier-envelope-phase effects}},}\
  }\href {\doibase 10.1103/PhysRevD.93.053011} {\bibfield  {journal} {\bibinfo
  {journal} {Phys. Rev. D}\ }\textbf {\bibinfo {volume} {93}},\ \bibinfo
  {pages} {053011} (\bibinfo {year} {2016})}\BibitemShut {NoStop}%
\bibitem [{\citenamefont {Nousch}\ \emph {et~al.}(2016)\citenamefont {Nousch},
  \citenamefont {Seipt}, \citenamefont {K\"ampfer},\ and\ \citenamefont
  {Titov}}]{NOUSCH2016162}%
  \BibitemOpen
  \bibfield  {author} {\bibinfo {author} {\bibfnamefont {T.}~\bibnamefont
  {Nousch}}, \bibinfo {author} {\bibfnamefont {D.}~\bibnamefont {Seipt}},
  \bibinfo {author} {\bibfnamefont {B.}~\bibnamefont {K\"ampfer}}, \ and\
  \bibinfo {author} {\bibfnamefont {A.~I.}\ \bibnamefont {Titov}},\ }\bibfield
  {title} {\enquote {\bibinfo {title} {{Spectral caustics in laser assisted
  Breit-Wheeler process}},}\ }\href {\doibase
  https://doi.org/10.1016/j.physletb.2016.01.062} {\bibfield  {journal}
  {\bibinfo  {journal} {Physics Letters B}\ }\textbf {\bibinfo {volume}
  {755}},\ \bibinfo {pages} {162 -- 167} (\bibinfo {year} {2016})}\BibitemShut
  {NoStop}%
\bibitem [{\citenamefont {Jansen}\ \emph {et~al.}(2016)\citenamefont {Jansen},
  \citenamefont {Kami\ifmmode~\acute{n}\else \'{n}\fi{}ski}, \citenamefont
  {Krajewska},\ and\ \citenamefont {M\"uller}}]{PhysRevD.94.013010}%
  \BibitemOpen
  \bibfield  {author} {\bibinfo {author} {\bibfnamefont {M.~J.~A.}\
  \bibnamefont {Jansen}}, \bibinfo {author} {\bibfnamefont {J.~Z.}\
  \bibnamefont {Kami\ifmmode~\acute{n}\else \'{n}\fi{}ski}}, \bibinfo {author}
  {\bibfnamefont {K.}~\bibnamefont {Krajewska}}, \ and\ \bibinfo {author}
  {\bibfnamefont {C.}~\bibnamefont {M\"uller}},\ }\bibfield  {title} {\enquote
  {\bibinfo {title} {{Strong-field Breit-Wheeler pair production in short laser
  pulses: Relevance of spin effects}},}\ }\href {\doibase
  10.1103/PhysRevD.94.013010} {\bibfield  {journal} {\bibinfo  {journal} {Phys.
  Rev. D}\ }\textbf {\bibinfo {volume} {94}},\ \bibinfo {pages} {013010}
  (\bibinfo {year} {2016})}\BibitemShut {NoStop}%
\bibitem [{\citenamefont {Meuren}\ \emph
  {et~al.}(2015{\natexlab{a}})\citenamefont {Meuren}, \citenamefont
  {Hatsagortsyan}, \citenamefont {Keitel},\ and\ \citenamefont
  {Di~Piazza}}]{PhysRevD.91.013009}%
  \BibitemOpen
  \bibfield  {author} {\bibinfo {author} {\bibfnamefont {S.}~\bibnamefont
  {Meuren}}, \bibinfo {author} {\bibfnamefont {K.~Z.}\ \bibnamefont
  {Hatsagortsyan}}, \bibinfo {author} {\bibfnamefont {C.~H.}\ \bibnamefont
  {Keitel}}, \ and\ \bibinfo {author} {\bibfnamefont {A.}~\bibnamefont
  {Di~Piazza}},\ }\bibfield  {title} {\enquote {\bibinfo {title}
  {Polarization-operator approach to pair creation in short laser pulses},}\
  }\href {\doibase 10.1103/PhysRevD.91.013009} {\bibfield  {journal} {\bibinfo
  {journal} {Phys. Rev. D}\ }\textbf {\bibinfo {volume} {91}},\ \bibinfo
  {pages} {013009} (\bibinfo {year} {2015}{\natexlab{a}})}\BibitemShut
  {NoStop}%
\bibitem [{\citenamefont {Meuren}\ \emph {et~al.}(2016)\citenamefont {Meuren},
  \citenamefont {Keitel},\ and\ \citenamefont
  {Di~Piazza}}]{PhysRevD.93.085028}%
  \BibitemOpen
  \bibfield  {author} {\bibinfo {author} {\bibfnamefont {S.}~\bibnamefont
  {Meuren}}, \bibinfo {author} {\bibfnamefont {C.~H.}\ \bibnamefont {Keitel}},
  \ and\ \bibinfo {author} {\bibfnamefont {A.}~\bibnamefont {Di~Piazza}},\
  }\bibfield  {title} {\enquote {\bibinfo {title} {Semiclassical picture for
  electron-positron photoproduction in strong laser fields},}\ }\href {\doibase
  10.1103/PhysRevD.93.085028} {\bibfield  {journal} {\bibinfo  {journal} {Phys.
  Rev. D}\ }\textbf {\bibinfo {volume} {93}},\ \bibinfo {pages} {085028}
  (\bibinfo {year} {2016})}\BibitemShut {NoStop}%
\bibitem [{\citenamefont {Meuren}\ \emph
  {et~al.}(2015{\natexlab{b}})\citenamefont {Meuren}, \citenamefont
  {Hatsagortsyan}, \citenamefont {Keitel},\ and\ \citenamefont
  {Di~Piazza}}]{PhysRevLett.114.143201}%
  \BibitemOpen
  \bibfield  {author} {\bibinfo {author} {\bibfnamefont {S.}~\bibnamefont
  {Meuren}}, \bibinfo {author} {\bibfnamefont {K.~Z.}\ \bibnamefont
  {Hatsagortsyan}}, \bibinfo {author} {\bibfnamefont {C.~H.}\ \bibnamefont
  {Keitel}}, \ and\ \bibinfo {author} {\bibfnamefont {A.}~\bibnamefont
  {Di~Piazza}},\ }\bibfield  {title} {\enquote {\bibinfo {title} {High-energy
  recollision processes of laser-generated electron-positron pairs},}\ }\href
  {\doibase 10.1103/PhysRevLett.114.143201} {\bibfield  {journal} {\bibinfo
  {journal} {Phys. Rev. Lett.}\ }\textbf {\bibinfo {volume} {114}},\ \bibinfo
  {pages} {143201} (\bibinfo {year} {2015}{\natexlab{b}})}\BibitemShut
  {NoStop}%
\bibitem [{\citenamefont {Baier}\ and\ \citenamefont
  {Katkov}(1968)}]{baier1968processes}%
  \BibitemOpen
  \bibfield  {author} {\bibinfo {author} {\bibfnamefont {V.N.}\ \bibnamefont
  {Baier}}\ and\ \bibinfo {author} {\bibfnamefont {V.M.}\ \bibnamefont
  {Katkov}},\ }\bibfield  {title} {\enquote {\bibinfo {title} {Processes
  involved in the motion of high energy particles in a magnetic field},}\
  }\href@noop {} {\bibfield  {journal} {\bibinfo  {journal} {J. Exp. Theor.
  Phys.}\ }\textbf {\bibinfo {volume} {26}},\ \bibinfo {pages} {854} (\bibinfo
  {year} {1968})}\BibitemShut {NoStop}%
\bibitem [{\citenamefont {Baier}\ \emph {et~al.}(1998)\citenamefont {Baier},
  \citenamefont {Katkov},\ and\ \citenamefont {Strakhovenko}}]{Baier1998}%
  \BibitemOpen
  \bibfield  {author} {\bibinfo {author} {\bibfnamefont {V.N.}\ \bibnamefont
  {Baier}}, \bibinfo {author} {\bibfnamefont {V.M.}\ \bibnamefont {Katkov}}, \
  and\ \bibinfo {author} {\bibfnamefont {V.M.}\ \bibnamefont {Strakhovenko}},\
  }\href@noop {} {\emph {\bibinfo {title} {{Electromagnetic Processes at High
  Energies in Oriented Single Crystals}}}}\ (\bibinfo  {publisher} {World
  Scientific},\ \bibinfo {year} {1998})\BibitemShut {NoStop}%
\bibitem [{\citenamefont {W\"ollert}\ \emph {et~al.}(2015)\citenamefont
  {W\"ollert}, \citenamefont {Bauke},\ and\ \citenamefont
  {Keitel}}]{PhysRevD.91.125026}%
  \BibitemOpen
  \bibfield  {author} {\bibinfo {author} {\bibfnamefont {A.}~\bibnamefont
  {W\"ollert}}, \bibinfo {author} {\bibfnamefont {H.}~\bibnamefont {Bauke}}, \
  and\ \bibinfo {author} {\bibfnamefont {C.~H.}\ \bibnamefont {Keitel}},\
  }\bibfield  {title} {\enquote {\bibinfo {title} {Spin polarized
  electron-positron pair production via elliptical polarized laser fields},}\
  }\href {\doibase 10.1103/PhysRevD.91.125026} {\bibfield  {journal} {\bibinfo
  {journal} {Phys. Rev. D}\ }\textbf {\bibinfo {volume} {91}},\ \bibinfo
  {pages} {125026} (\bibinfo {year} {2015})}\BibitemShut {NoStop}%
\bibitem [{\citenamefont {Chen}\ \emph {et~al.}(2019)\citenamefont {Chen},
  \citenamefont {He}, \citenamefont {Shaisultanov}, \citenamefont
  {Hatsagortsyan},\ and\ \citenamefont {Keitel}}]{PhysRevLett.123.174801}%
  \BibitemOpen
  \bibfield  {author} {\bibinfo {author} {\bibfnamefont {Y.-Y.}\ \bibnamefont
  {Chen}}, \bibinfo {author} {\bibfnamefont {P.-L.}\ \bibnamefont {He}},
  \bibinfo {author} {\bibfnamefont {R.}~\bibnamefont {Shaisultanov}}, \bibinfo
  {author} {\bibfnamefont {K.~Z.}\ \bibnamefont {Hatsagortsyan}}, \ and\
  \bibinfo {author} {\bibfnamefont {C.~H.}\ \bibnamefont {Keitel}},\ }\bibfield
   {title} {\enquote {\bibinfo {title} {Polarized positron beams via intense
  two-color laser pulses},}\ }\href {\doibase 10.1103/PhysRevLett.123.174801}
  {\bibfield  {journal} {\bibinfo  {journal} {Phys. Rev. Lett.}\ }\textbf
  {\bibinfo {volume} {123}},\ \bibinfo {pages} {174801} (\bibinfo {year}
  {2019})}\BibitemShut {NoStop}%
\bibitem [{\citenamefont {Wan}\ \emph {et~al.}(2020)\citenamefont {Wan},
  \citenamefont {Shaisultanov}, \citenamefont {Li}, \citenamefont
  {Hatsagortsyan}, \citenamefont {Keitel},\ and\ \citenamefont
  {Li}}]{WAN2020135120}%
  \BibitemOpen
  \bibfield  {author} {\bibinfo {author} {\bibfnamefont {F.}~\bibnamefont
  {Wan}}, \bibinfo {author} {\bibfnamefont {R.}~\bibnamefont {Shaisultanov}},
  \bibinfo {author} {\bibfnamefont {Y.-F.}\ \bibnamefont {Li}}, \bibinfo
  {author} {\bibfnamefont {K.~Z.}\ \bibnamefont {Hatsagortsyan}}, \bibinfo
  {author} {\bibfnamefont {C.~H.}\ \bibnamefont {Keitel}}, \ and\ \bibinfo
  {author} {\bibfnamefont {J.-X.}\ \bibnamefont {Li}},\ }\bibfield  {title}
  {\enquote {\bibinfo {title} {Ultrarelativistic polarized positron jets via
  collision of electron and ultraintense laser beams},}\ }\href {\doibase
  https://doi.org/10.1016/j.physletb.2019.135120} {\bibfield  {journal}
  {\bibinfo  {journal} {Physics Letters B}\ }\textbf {\bibinfo {volume}
  {800}},\ \bibinfo {pages} {135120} (\bibinfo {year} {2020})}\BibitemShut
  {NoStop}%
\bibitem [{\citenamefont {Wistisen}\ and\ \citenamefont
  {Di~Piazza}(2019{\natexlab{a}})}]{Wistisen2019}%
  \BibitemOpen
  \bibfield  {author} {\bibinfo {author} {\bibfnamefont {T.~N.}\ \bibnamefont
  {Wistisen}}\ and\ \bibinfo {author} {\bibfnamefont {A.}~\bibnamefont
  {Di~Piazza}},\ }\bibfield  {title} {\enquote {\bibinfo {title} {Complete
  treatment of single-photon emission in planar channeling},}\ }\href {\doibase
  10.1103/PhysRevD.99.116010} {\bibfield  {journal} {\bibinfo  {journal} {Phys.
  Rev. D}\ }\textbf {\bibinfo {volume} {99}},\ \bibinfo {pages} {116010}
  (\bibinfo {year} {2019}{\natexlab{a}})}\BibitemShut {NoStop}%
\bibitem [{\citenamefont {Wistisen}\ and\ \citenamefont
  {Di~Piazza}(2018)}]{PhysRevA.98.022131}%
  \BibitemOpen
  \bibfield  {author} {\bibinfo {author} {\bibfnamefont {T.~N.}\ \bibnamefont
  {Wistisen}}\ and\ \bibinfo {author} {\bibfnamefont {A.}~\bibnamefont
  {Di~Piazza}},\ }\bibfield  {title} {\enquote {\bibinfo {title} {{Impact of
  the quantized transverse motion on radiation emission in a Dirac harmonic
  oscillator}},}\ }\href {\doibase 10.1103/PhysRevA.98.022131} {\bibfield
  {journal} {\bibinfo  {journal} {Phys. Rev. A}\ }\textbf {\bibinfo {volume}
  {98}},\ \bibinfo {pages} {022131} (\bibinfo {year} {2018})}\BibitemShut
  {NoStop}%
\bibitem [{\citenamefont {Wistisen}\ and\ \citenamefont
  {Di~Piazza}(2019{\natexlab{b}})}]{PhysRevD.100.116001}%
  \BibitemOpen
  \bibfield  {author} {\bibinfo {author} {\bibfnamefont {T.~N.}\ \bibnamefont
  {Wistisen}}\ and\ \bibinfo {author} {\bibfnamefont {A.}~\bibnamefont
  {Di~Piazza}},\ }\bibfield  {title} {\enquote {\bibinfo {title} {Numerical
  approach to the semiclassical method of radiation emission for arbitrary
  electron spin and photon polarization},}\ }\href {\doibase
  10.1103/PhysRevD.100.116001} {\bibfield  {journal} {\bibinfo  {journal}
  {Phys. Rev. D}\ }\textbf {\bibinfo {volume} {100}},\ \bibinfo {pages}
  {116001} (\bibinfo {year} {2019}{\natexlab{b}})}\BibitemShut {NoStop}%
\bibitem [{\citenamefont {Wistisen}(2014)}]{PhysRevD.90.125008}%
  \BibitemOpen
  \bibfield  {author} {\bibinfo {author} {\bibfnamefont {T.~N.}\ \bibnamefont
  {Wistisen}},\ }\bibfield  {title} {\enquote {\bibinfo {title} {{Interference
  effect in nonlinear Compton scattering}},}\ }\href {\doibase
  10.1103/PhysRevD.90.125008} {\bibfield  {journal} {\bibinfo  {journal} {Phys.
  Rev. D}\ }\textbf {\bibinfo {volume} {90}},\ \bibinfo {pages} {125008}
  (\bibinfo {year} {2014})}\BibitemShut {NoStop}%
\bibitem [{\citenamefont {Wistisen}(2015)}]{PhysRevD.92.045045}%
  \BibitemOpen
  \bibfield  {author} {\bibinfo {author} {\bibfnamefont {T.~N.}\ \bibnamefont
  {Wistisen}},\ }\bibfield  {title} {\enquote {\bibinfo {title} {Quantum
  synchrotron radiation in the case of a field with finite extension},}\ }\href
  {\doibase 10.1103/PhysRevD.92.045045} {\bibfield  {journal} {\bibinfo
  {journal} {Phys. Rev. D}\ }\textbf {\bibinfo {volume} {92}},\ \bibinfo
  {pages} {045045} (\bibinfo {year} {2015})}\BibitemShut {NoStop}%
\bibitem [{\citenamefont {Ritus}(1985)}]{Ritus}%
  \BibitemOpen
  \bibfield  {author} {\bibinfo {author} {\bibfnamefont {V.I.}\ \bibnamefont
  {Ritus}},\ }\bibfield  {title} {{\selectlanguage {English}\enquote {\bibinfo
  {title} {Quantum effects of the interaction of elementary particles with an
  intense electromagnetic field},}\ }}\href {\doibase 10.1007/BF01120220}
  {\bibfield  {journal} {\bibinfo  {journal} {Journal of Soviet Laser
  Research}\ }\textbf {\bibinfo {volume} {6}},\ \bibinfo {pages} {497--617}
  (\bibinfo {year} {1985})}\BibitemShut {NoStop}%
\bibitem [{\citenamefont {Hartin}\ \emph {et~al.}(2019)\citenamefont {Hartin},
  \citenamefont {Ringwald},\ and\ \citenamefont {Tapia}}]{PhysRevD.99.036008}%
  \BibitemOpen
  \bibfield  {author} {\bibinfo {author} {\bibfnamefont {A.}~\bibnamefont
  {Hartin}}, \bibinfo {author} {\bibfnamefont {A.}~\bibnamefont {Ringwald}}, \
  and\ \bibinfo {author} {\bibfnamefont {N.}~\bibnamefont {Tapia}},\ }\bibfield
   {title} {\enquote {\bibinfo {title} {Measuring the boiling point of the
  vacuum of quantum electrodynamics},}\ }\href {\doibase
  10.1103/PhysRevD.99.036008} {\bibfield  {journal} {\bibinfo  {journal} {Phys.
  Rev. D}\ }\textbf {\bibinfo {volume} {99}},\ \bibinfo {pages} {036008}
  (\bibinfo {year} {2019})}\BibitemShut {NoStop}%
\bibitem [{\citenamefont {Heinzl}\ \emph {et~al.}(2006)\citenamefont {Heinzl},
  \citenamefont {Liesfeld}, \citenamefont {Amthor}, \citenamefont {Schwoerer},
  \citenamefont {Sauerbrey},\ and\ \citenamefont {Wipf}}]{Heinzl2006a}%
  \BibitemOpen
  \bibfield  {author} {\bibinfo {author} {\bibfnamefont {T.}~\bibnamefont
  {Heinzl}}, \bibinfo {author} {\bibfnamefont {B.}~\bibnamefont {Liesfeld}},
  \bibinfo {author} {\bibfnamefont {K.U.}\ \bibnamefont {Amthor}}, \bibinfo
  {author} {\bibfnamefont {H.}~\bibnamefont {Schwoerer}}, \bibinfo {author}
  {\bibfnamefont {R.}~\bibnamefont {Sauerbrey}}, \ and\ \bibinfo {author}
  {\bibfnamefont {A.}~\bibnamefont {Wipf}},\ }\bibfield  {title} {\enquote
  {\bibinfo {title} {On the observation of vacuum birefringence},}\ }\href
  {\doibase http://dx.doi.org/10.1016/j.optcom.2006.06.053} {\bibfield
  {journal} {\bibinfo  {journal} {Optics Communications}\ }\textbf {\bibinfo
  {volume} {267}},\ \bibinfo {pages} {318 -- 321} (\bibinfo {year}
  {2006})}\BibitemShut {NoStop}%
\bibitem [{\citenamefont {Di~Piazza}\ \emph {et~al.}(2006)\citenamefont
  {Di~Piazza}, \citenamefont {Hatsagortsyan},\ and\ \citenamefont
  {Keitel}}]{PhysRevLett.97.083603}%
  \BibitemOpen
  \bibfield  {author} {\bibinfo {author} {\bibfnamefont {A.}~\bibnamefont
  {Di~Piazza}}, \bibinfo {author} {\bibfnamefont {K.~Z.}\ \bibnamefont
  {Hatsagortsyan}}, \ and\ \bibinfo {author} {\bibfnamefont {C.~H.}\
  \bibnamefont {Keitel}},\ }\bibfield  {title} {\enquote {\bibinfo {title}
  {Light diffraction by a strong standing electromagnetic wave},}\ }\href
  {\doibase 10.1103/PhysRevLett.97.083603} {\bibfield  {journal} {\bibinfo
  {journal} {Phys. Rev. Lett.}\ }\textbf {\bibinfo {volume} {97}},\ \bibinfo
  {pages} {083603} (\bibinfo {year} {2006})}\BibitemShut {NoStop}%
\bibitem [{\citenamefont {Adler}(2007)}]{1751-8121-40-5-F01}%
  \BibitemOpen
  \bibfield  {author} {\bibinfo {author} {\bibfnamefont {S.~L.}\ \bibnamefont
  {Adler}},\ }\bibfield  {title} {\enquote {\bibinfo {title} {Vacuum
  birefringence in a rotating magnetic field},}\ }\href
  {http://stacks.iop.org/1751-8121/40/i=5/a=F01} {\bibfield  {journal}
  {\bibinfo  {journal} {Journal of Physics A: Mathematical and Theoretical}\
  }\textbf {\bibinfo {volume} {40}},\ \bibinfo {pages} {F143} (\bibinfo {year}
  {2007})}\BibitemShut {NoStop}%
\bibitem [{\citenamefont {Karbstein}\ \emph {et~al.}(2015)\citenamefont
  {Karbstein}, \citenamefont {Gies}, \citenamefont {Reuter},\ and\
  \citenamefont {Zepf}}]{PhysRevD.92.071301}%
  \BibitemOpen
  \bibfield  {author} {\bibinfo {author} {\bibfnamefont {F.}~\bibnamefont
  {Karbstein}}, \bibinfo {author} {\bibfnamefont {H.}~\bibnamefont {Gies}},
  \bibinfo {author} {\bibfnamefont {M.}~\bibnamefont {Reuter}}, \ and\ \bibinfo
  {author} {\bibfnamefont {M.}~\bibnamefont {Zepf}},\ }\bibfield  {title}
  {\enquote {\bibinfo {title} {Vacuum birefringence in strong inhomogeneous
  electromagnetic fields},}\ }\href {\doibase 10.1103/PhysRevD.92.071301}
  {\bibfield  {journal} {\bibinfo  {journal} {Phys. Rev. D}\ }\textbf {\bibinfo
  {volume} {92}},\ \bibinfo {pages} {071301} (\bibinfo {year}
  {2015})}\BibitemShut {NoStop}%
\bibitem [{\citenamefont {Bregant}\ \emph {et~al.}(2008)\citenamefont
  {Bregant}, \citenamefont {Cantatore}, \citenamefont {Carusotto},
  \citenamefont {Cimino}, \citenamefont {Della~Valle}, \citenamefont
  {Di~Domenico}, \citenamefont {Gastaldi}, \citenamefont {Karuza},
  \citenamefont {Lozza}, \citenamefont {Milotti}, \citenamefont {Polacco},
  \citenamefont {Raiteri}, \citenamefont {Ruoso}, \citenamefont {Zavattini},\
  and\ \citenamefont {Zavattini}}]{PhysRevD.78.032006}%
  \BibitemOpen
  \bibfield  {author} {\bibinfo {author} {\bibfnamefont {M.}~\bibnamefont
  {Bregant}}, \bibinfo {author} {\bibfnamefont {G.}~\bibnamefont {Cantatore}},
  \bibinfo {author} {\bibfnamefont {S.}~\bibnamefont {Carusotto}}, \bibinfo
  {author} {\bibfnamefont {R.}~\bibnamefont {Cimino}}, \bibinfo {author}
  {\bibfnamefont {F.}~\bibnamefont {Della~Valle}}, \bibinfo {author}
  {\bibfnamefont {G.}~\bibnamefont {Di~Domenico}}, \bibinfo {author}
  {\bibfnamefont {U.}~\bibnamefont {Gastaldi}}, \bibinfo {author}
  {\bibfnamefont {M.}~\bibnamefont {Karuza}}, \bibinfo {author} {\bibfnamefont
  {V.}~\bibnamefont {Lozza}}, \bibinfo {author} {\bibfnamefont
  {E.}~\bibnamefont {Milotti}}, \bibinfo {author} {\bibfnamefont
  {E.}~\bibnamefont {Polacco}}, \bibinfo {author} {\bibfnamefont
  {G.}~\bibnamefont {Raiteri}}, \bibinfo {author} {\bibfnamefont
  {G.}~\bibnamefont {Ruoso}}, \bibinfo {author} {\bibfnamefont
  {E.}~\bibnamefont {Zavattini}}, \ and\ \bibinfo {author} {\bibfnamefont
  {G.}~\bibnamefont {Zavattini}} (\bibinfo {collaboration} {PVLAS
  Collaboration}),\ }\bibfield  {title} {\enquote {\bibinfo {title} {Limits on
  low energy photon-photon scattering from an experiment on magnetic vacuum
  birefringence},}\ }\href {\doibase 10.1103/PhysRevD.78.032006} {\bibfield
  {journal} {\bibinfo  {journal} {Phys. Rev. D}\ }\textbf {\bibinfo {volume}
  {78}},\ \bibinfo {pages} {032006} (\bibinfo {year} {2008})}\BibitemShut
  {NoStop}%
\bibitem [{\citenamefont {Bakalov}\ \emph {et~al.}(1994)\citenamefont
  {Bakalov}, \citenamefont {Cantatore}, \citenamefont {Carugno}, \citenamefont
  {Carusotto}, \citenamefont {Favaron}, \citenamefont {Valle}, \citenamefont
  {Gabrielli}, \citenamefont {Gastaldi}, \citenamefont {Iacopini},
  \citenamefont {Micossi}, \citenamefont {Milotti}, \citenamefont {Onofrio},
  \citenamefont {Pengo}, \citenamefont {Perrone}, \citenamefont {Petrucci},
  \citenamefont {Polacco}, \citenamefont {Rizzo}, \citenamefont {Ruoso},
  \citenamefont {Zavattini},\ and\ \citenamefont {Zavattini}}]{BAKALOV1994180}%
  \BibitemOpen
  \bibfield  {author} {\bibinfo {author} {\bibfnamefont {D.}~\bibnamefont
  {Bakalov}}, \bibinfo {author} {\bibfnamefont {G.}~\bibnamefont {Cantatore}},
  \bibinfo {author} {\bibfnamefont {G.}~\bibnamefont {Carugno}}, \bibinfo
  {author} {\bibfnamefont {S.}~\bibnamefont {Carusotto}}, \bibinfo {author}
  {\bibfnamefont {P.}~\bibnamefont {Favaron}}, \bibinfo {author} {\bibfnamefont
  {F.~Della}\ \bibnamefont {Valle}}, \bibinfo {author} {\bibfnamefont
  {I.}~\bibnamefont {Gabrielli}}, \bibinfo {author} {\bibfnamefont
  {U.}~\bibnamefont {Gastaldi}}, \bibinfo {author} {\bibfnamefont
  {E.}~\bibnamefont {Iacopini}}, \bibinfo {author} {\bibfnamefont
  {P.}~\bibnamefont {Micossi}}, \bibinfo {author} {\bibfnamefont
  {E.}~\bibnamefont {Milotti}}, \bibinfo {author} {\bibfnamefont
  {R.}~\bibnamefont {Onofrio}}, \bibinfo {author} {\bibfnamefont
  {R.}~\bibnamefont {Pengo}}, \bibinfo {author} {\bibfnamefont
  {F.}~\bibnamefont {Perrone}}, \bibinfo {author} {\bibfnamefont
  {G.}~\bibnamefont {Petrucci}}, \bibinfo {author} {\bibfnamefont
  {E.}~\bibnamefont {Polacco}}, \bibinfo {author} {\bibfnamefont
  {C.}~\bibnamefont {Rizzo}}, \bibinfo {author} {\bibfnamefont
  {G.}~\bibnamefont {Ruoso}}, \bibinfo {author} {\bibfnamefont
  {E.}~\bibnamefont {Zavattini}}, \ and\ \bibinfo {author} {\bibfnamefont
  {G.}~\bibnamefont {Zavattini}},\ }\bibfield  {title} {\enquote {\bibinfo
  {title} {{PVLAS: Vacuum Birefringence and production and detection of nearly
  massless, weakly coupled particles by optical techniques.}}}\ }\href
  {\doibase https://doi.org/10.1016/0920-5632(94)90243-7} {\bibfield  {journal}
  {\bibinfo  {journal} {Nuclear Physics B - Proceedings Supplements}\ }\textbf
  {\bibinfo {volume} {35}},\ \bibinfo {pages} {180 -- 182} (\bibinfo {year}
  {1994})}\BibitemShut {NoStop}%
\bibitem [{\citenamefont {King}\ and\ \citenamefont
  {Elkina}(2016)}]{PhysRevA.94.062102}%
  \BibitemOpen
  \bibfield  {author} {\bibinfo {author} {\bibfnamefont {B.}~\bibnamefont
  {King}}\ and\ \bibinfo {author} {\bibfnamefont {N.}~\bibnamefont {Elkina}},\
  }\bibfield  {title} {\enquote {\bibinfo {title} {Vacuum birefringence in
  high-energy laser-electron collisions},}\ }\href {\doibase
  10.1103/PhysRevA.94.062102} {\bibfield  {journal} {\bibinfo  {journal} {Phys.
  Rev. A}\ }\textbf {\bibinfo {volume} {94}},\ \bibinfo {pages} {062102}
  (\bibinfo {year} {2016})}\BibitemShut {NoStop}%
\bibitem [{\citenamefont {Wistisen}\ and\ \citenamefont
  {Uggerh\o{}j}(2013)}]{PhysRevD.88.053009}%
  \BibitemOpen
  \bibfield  {author} {\bibinfo {author} {\bibfnamefont {T.~N.}\ \bibnamefont
  {Wistisen}}\ and\ \bibinfo {author} {\bibfnamefont {U.~I.}\ \bibnamefont
  {Uggerh\o{}j}},\ }\bibfield  {title} {\enquote {\bibinfo {title} {{Vacuum
  birefringence by Compton backscattering through a strong field}},}\ }\href
  {\doibase 10.1103/PhysRevD.88.053009} {\bibfield  {journal} {\bibinfo
  {journal} {Phys. Rev. D}\ }\textbf {\bibinfo {volume} {88}},\ \bibinfo
  {pages} {053009} (\bibinfo {year} {2013})}\BibitemShut {NoStop}%
\bibitem [{\citenamefont {Bragin}\ \emph {et~al.}(2017)\citenamefont {Bragin},
  \citenamefont {Meuren}, \citenamefont {Keitel},\ and\ \citenamefont
  {Di~Piazza}}]{PhysRevLett.119.250403}%
  \BibitemOpen
  \bibfield  {author} {\bibinfo {author} {\bibfnamefont {S.}~\bibnamefont
  {Bragin}}, \bibinfo {author} {\bibfnamefont {S.}~\bibnamefont {Meuren}},
  \bibinfo {author} {\bibfnamefont {C.~H.}\ \bibnamefont {Keitel}}, \ and\
  \bibinfo {author} {\bibfnamefont {A.}~\bibnamefont {Di~Piazza}},\ }\bibfield
  {title} {\enquote {\bibinfo {title} {High-energy vacuum birefringence and
  dichroism in an ultrastrong laser field},}\ }\href {\doibase
  10.1103/PhysRevLett.119.250403} {\bibfield  {journal} {\bibinfo  {journal}
  {Phys. Rev. Lett.}\ }\textbf {\bibinfo {volume} {119}},\ \bibinfo {pages}
  {250403} (\bibinfo {year} {2017})}\BibitemShut {NoStop}%
\bibitem [{\citenamefont {Denisov}\ \emph {et~al.}(2017)\citenamefont
  {Denisov}, \citenamefont {Dolgaya},\ and\ \citenamefont
  {Sokolov}}]{Denisov2017}%
  \BibitemOpen
  \bibfield  {author} {\bibinfo {author} {\bibfnamefont {V.I.}\ \bibnamefont
  {Denisov}}, \bibinfo {author} {\bibfnamefont {E.E.}\ \bibnamefont {Dolgaya}},
  \ and\ \bibinfo {author} {\bibfnamefont {V.A.}\ \bibnamefont {Sokolov}},\
  }\bibfield  {title} {\enquote {\bibinfo {title} {{Nonperturbative QED vacuum
  birefringence}},}\ }\href {\doibase 10.1007/JHEP05(2017)105} {\bibfield
  {journal} {\bibinfo  {journal} {Journal of High Energy Physics}\ }\textbf
  {\bibinfo {volume} {2017}},\ \bibinfo {pages} {105} (\bibinfo {year}
  {2017})}\BibitemShut {NoStop}%
\bibitem [{\citenamefont {Karbstein}(2018)}]{PhysRevD.98.056010}%
  \BibitemOpen
  \bibfield  {author} {\bibinfo {author} {\bibfnamefont {F.}~\bibnamefont
  {Karbstein}},\ }\bibfield  {title} {\enquote {\bibinfo {title} {Vacuum
  birefringence in the head-on collision of x-ray free-electron laser and
  optical high-intensity laser pulses},}\ }\href {\doibase
  10.1103/PhysRevD.98.056010} {\bibfield  {journal} {\bibinfo  {journal} {Phys.
  Rev. D}\ }\textbf {\bibinfo {volume} {98}},\ \bibinfo {pages} {056010}
  (\bibinfo {year} {2018})}\BibitemShut {NoStop}%
\bibitem [{\citenamefont {Briscese}(2018)}]{PhysRevA.97.033803}%
  \BibitemOpen
  \bibfield  {author} {\bibinfo {author} {\bibfnamefont {F.}~\bibnamefont
  {Briscese}},\ }\bibfield  {title} {\enquote {\bibinfo {title} {Collective
  behavior of light in vacuum},}\ }\href {\doibase 10.1103/PhysRevA.97.033803}
  {\bibfield  {journal} {\bibinfo  {journal} {Phys. Rev. A}\ }\textbf {\bibinfo
  {volume} {97}},\ \bibinfo {pages} {033803} (\bibinfo {year}
  {2018})}\BibitemShut {NoStop}%
\bibitem [{\citenamefont {Ter-Mikaelian}(1972)}]{Ter-Mikaelian1972}%
  \BibitemOpen
  \bibfield  {author} {\bibinfo {author} {\bibfnamefont {M.~L.}\ \bibnamefont
  {Ter-Mikaelian}},\ }\href@noop {} {{\selectlanguage {English}\emph {\bibinfo
  {title} {High-energy electromagnetic processes in condensed media}}}}\
  (\bibinfo  {publisher} {Wiley-Interscience},\ \bibinfo {year}
  {1972})\BibitemShut {NoStop}%
\end{thebibliography}%

\end{document}